\documentclass[journal]{IEEEtran}
\usepackage{amsmath,graphicx}
\usepackage{amssymb}  
\usepackage[utf8]{inputenc}
\usepackage[amsmath,thmmarks,hyperref]{ntheorem} 
\usepackage{bbm}               
\usepackage{mathtools}         
\usepackage{stmaryrd}          
\usepackage{paralist}          
\usepackage{mathrsfs}
\usepackage{hyperref}
\usepackage{textcomp}
\usepackage{color}             
\usepackage{multicol}
\usepackage{booktabs}
\usepackage{subcaption}
\renewcommand{\arraystretch}{1.2} 
\usepackage[export]{adjustbox} 
\usepackage[normalem]{ulem} 
\usepackage{tikz}
\usetikzlibrary{arrows}
\usetikzlibrary{matrix}

\usepackage{algorithm}
\usepackage{algpseudocode}

\graphicspath{{pics/}}

\newcommand{\mypar}[1]{{\bf #1.}}
\newcommand{\coord}[1]{\text{\bf #1}}

\newcommand{\R}[0]{{\mathbb{R}}}

\newcommand{\diag}[0]{\operatorname{diag}}

\newcommand{\one}[0]{I}

\newcommand{\ra}[1]{\renewcommand{\arraystretch}{#1}}
\newcommand{\cl}[1]{\overline{#1}}

\newcommand{\meet}[0]{\wedge}

\newcommand{\ft}[1]{\widehat{#1}}
\newcommand{\fr}[1]{{#1}'}
\newcommand{\chr}[1]{\iota_{\{#1\}}}
\newcommand{\TV}[0]{\operatorname{TV}}
\newcommand{\STV}[0]{\operatorname{STV}}
\newcommand{\dg}[0]{{\cal D}}
\newcommand{\vrt}[0]{{\cal V}}
\newcommand{\pos}[0]{{\cal P}}
\newcommand{\edg}[0]{{\cal E}}
\newcommand{\graph}[0]{{\cal G}}

\newcommand{\E}{e}

\DeclarePairedDelimiter\ideal\langle\rangle
\reDeclarePairedDelimiterInnerWrapper\ideal{star}{
\mathopen{#1\vphantom{\MTkillspecial{#2}}\kern-\nulldelimiterspace\right.}
#2
\mathclose{\left.\kern-\nulldelimiterspace\vphantom{\MTkillspecial{#2}}#3}}

\DeclarePairedDelimiter{\norm}{\lVert}{\rVert}

\newcommand{\Fourier}{F}

\newcommand{\prob}{\operatorname{prob}}

\newtheorem{theorem}{Theorem}
\newtheorem{definition}{Definition}

\theoremheaderfont{\scshape}
\theorembodyfont{\normalfont}
\theoremstyle{nonumberplain}
\theoremseparator{:}
\theoremsymbol{\hbox{$\boxempty$}}

\title{Causal Fourier Analysis on Directed Acyclic Graphs and Posets}

\author{Bastian~Seifert,~\IEEEmembership{Member,~IEEE}
  Chris~Wendler,~\IEEEmembership{Student Member,~IEEE}
  Markus~Püschel,~\IEEEmembership{Fellow,~IEEE}%
  \thanks{The authors are with the Department of Computer Science, ETH
    Zurich, Switzerland (email: seifert.bastian@protonmail.com,
    chris.wendler@inf.ethz.ch, pueschel@inf.ethz.ch )}
  \thanks{Manuscript received ???; revised ???}}
    
\begin{document}

\maketitle

\begin{abstract}
We present a novel form of Fourier analysis, and associated signal processing concepts, for signals (or data) indexed by edge-weighted directed acyclic graphs (DAGs). This means that our Fourier basis yields an eigendecomposition of a suitable notion of shift and convolution operators that we define. DAGs are the common model to capture causal relationships between data values and in this case our proposed Fourier analysis relates data with its causes under a linearity assumption that we define. The definition of the Fourier transform requires the transitive closure of the weighted DAG for which several forms are possible depending on the interpretation of the edge weights. Examples include level of influence, distance, or pollution distribution. Our framework is specific to DAGs and leverages, and extends, the classical theory of Moebius inversion from combinatorics. For a prototypical application we consider the reconstruction of signals from samples assuming Fourier-sparsity, i.e., few causes. In particular, we consider DAGs modeling dynamic networks in which edges change over time. We model the spread of an infection on such a DAG obtained from real-world contact tracing data and learn the infection signal from samples.
\end{abstract}

\begin{IEEEkeywords}
	Graph signal processing, DAG, partial order, causality, structural equation model, Moebius inversion, Fourier transform, convolution, non-Euclidean, Fourier sparsity, dynamic graph, infection spreading, binary classifier
 \end{IEEEkeywords}

\section{Introduction}\label{sec:introduction}%

Causality studies which events influence others building on powerful classical theories including Bayesian networks and structural causal models~\cite{Koller:2009,Peters:2017}. However, understanding and deriving causality from data continues to be a challenging problem in data science and machine learning~\cite{Schoelkopf:2022a}. The common index domains for causal data are directed acyclic graphs (DAGs), in which the nodes represent events and the directed edges causal relationships. Motivated by their importance, we propose a novel form of Fourier analysis for signals (or data) on DAGs, including associated signal processing (SP) concepts of shift, convolution, spectrum, frequency response, and others. DAGs are closely related to partially ordered sets (posets), where the partial order determines whether a node is a predecessor of another node. Thus, our SP framework can equivalently be considered for signals on posets.

Our framework is specific for DAGs and fundamentally different from prior graph SP based on Laplacian or adjacency matrix~\cite{Shuman:13,Sandryhaila:13}, which fails for DAGs due to a collapsing spectrum. The causal nature of our framework is reflected in both shift and associated Fourier transform as will become clear later. Before we state our contribution in greater detail we provide the context of prior work. A more detailed discussion of related work is provided later in Section~\ref{sec:dis}.

\mypar{Graph signal processing} Prior graph SP uses the eigenbasis of adjacency matrix or Laplacian as Fourier basis and to define related concepts~\cite{Sandryhaila:13,Shuman:13}. For undirected graphs both exist and are even orthogonal. For directed graphs (digraphs) this is not the case, the more general Jordan normal form is not computable, and thus a proper digraph SP was still considered an open problem in~\cite[Sec.~III.A]{Ortega.Frossard.Kovacevic.Moura.Vandergheynst:2018a} despite various applications~\cite{Marques.Segarra.Mateos:2020a}. Several solutions have been proposed, mostly based on a form of approximation or relaxation of requirements~\cite{Sardellitti.Barbarossa.DiLorenzo:2017a,Shafipour.Khodabakhsh.Mateos.Nikolova:2018a,Shafipour.Khodabakhsh.Mateos.Nikolova:2019a,Furutani.Shibahara.Akiyama.Hato.Aida:2019a,Domingos.Moura:2020a,Seifert.Pueschel:2020a}.

DAGs constitute a worst case in digraph SP since they are associated with triangular Laplacian or adjacency matrices. In particular the latter have only one eigenvalue zero and thus never an eigenbasis. The lack of a proper form of DAG Fourier analysis prevents the application of SP methods to causal data indexed by DAGs.

\mypar{Causality} Classical models for causality include Bayesian networks \cite{Koller:2009}, which encode multivariate probability distributions and enable different forms of causal reasoning. Structural causal models (SCMs), also called structural equation models (SEMs) \cite{Peters:2017}, define how variables associated with DAG nodes are computed from parent nodes. Both approaches are probabilistic and model data as random vectors on DAGs, in which the edges represent causal dependencies. Despite powerful theory, learning causality from data is hard with many pitfalls~\cite{Vowels.Camgoz.Bowden:2021a,Guo.Cheng.Li.Hahn.Liu:2020a,Schoelkopf:2022a}. In particular, causal data typically has a DAG as index domain, but the converse does not hold: if data is given on a DAG, its edges do not generally imply causal relationships due to possible hidden confounding variables, and detecting causality does require additional techniques such as interventions \cite{Peters:2017}. One important line of work in causal reasoning addresses the problem of learning the DAG from observed data~\cite{zheng2018notears,ng2020GOLEM}.

Our proposed Fourier analysis could bring a novel view point and SP-inspired tools to the analysis of causal data. As a first step, we  have tackled a novel variant of DAG learning~\cite{misiakos2023learning}, based on the assumption of sparsity in the Fourier domain.

\mypar{Contribution} We present a novel form of Fourier analysis, and associated basic SP concepts, for signals (or data) indexed by the nodes of a weighted DAG, extending and completing our preliminary work in~\cite{Seifert.Wendler.Pueschel:2022a}. Our framework can be used on any DAG signal, whether the DAG captures causality or not. However, in the causal case, and under assumptions and in a sense that we define, the causes of a signal become its spectrum on the DAG in our Fourier analysis. Further, our framework can be related to the special class of linear SEMs, providing a novel Fourier-perspective.

In contrast to prior graph SP, our work leverages the partial order structure defined by a DAG, which also makes it specific to the acyclic case. The weighted DAG describes how the signal value at a node is determined (or caused in the causal case) by its parent nodes. But, by transitivity, this means that the value is determined by all predecessors. We assume this relation to be linear and thus it is obtained by a suitable form of weighted transitive closure of the DAG~\cite{Lehmann:1977a}, whose form depends on the meaning of the weights, such as distance, level of influence, or fraction of propagation. Viewed as a matrix, the transitive closure determines the Fourier basis relating a signal to its causes, which become its spectrum. We define an associated notion of shift that operates in the frequency domain by removing causes, and show that the spectrum is partially ordered isomorphic to the DAG.

Our prototypical experiments consider the reconstruction of DAG signals from samples under the assumption of sparsity in the Fourier domain. We show a synthetic experiment as proof of concept. Then we consider one possible application domain for our work: dynamic networks whose edges change over time, which can be modeled as DAGs by unrolling the time dimension and connecting subsequent iterations of the graphs accordingly. We model the spread of a disease on such a DAG, derived from real-world contact tracing data. Then we learn the infection signal from samples. Our causal Fourier basis yields superior results when compared to prior graph Fourier bases, which require dropping the directionality of the edges.

\section{DAGs and Posets}
\label{sec:DAGsAndPosets}%

We explain the necessary background on directed acyclic graphs (DAGs), partially ordered sets (posets), and their close relationship. Sets and graphs are denoted with calligraphic letters, matrices in upper case, vectors in bold lower case, and scalars in lower case. 

\mypar{DAGs} A directed graph (digraph) $\dg = (\vrt,\edg)$ consists of a finite set $\vrt$ of $n$ nodes and a set $\edg$ of $m$ directed edges: $\edg \subseteq \{ (y,x) \; | \; x,y \in \vrt\}$. $\dg$ is acyclic, and thus a DAG, if it contains no cycles. Since $\dg$ is acyclic, we can sort $\vrt$ topologically, which means $(y,x)\in \edg$ implies that $x$ comes after $y$. We consider weighted DAGs $(\vrt,\edg,A)$, in which each edge $(y,x)$ is assigned a nonzero (not necessarily positive) weight $a_{x,y}$. These are collected in the matrix
\begin{equation}
	\label{eq:AdjacencyMatrixDAG}
	A = (a_{x,y})_{x,y \in\vrt} =
	\begin{cases}
		a_{x,y} & \text{if } (y,x) \in E, \\
		0 & \text{otherwise}.
	\end{cases}
\end{equation}
The topological sort makes $A$ lower triangular with zeros on the diagonal. If all weights are $=1$, $A$ is just the adjacency matrix.

\mypar{Posets} A partially ordered set (poset) \cite{Stanley:2011} is a finite set $\pos$ with a partial order, i.e., a binary relation $\leq$ that satisfies for all $x,y,z \in\pos$
\begin{enumerate}
    \item $x \leq x$ (reflexivity),
    \item $y \leq x$ and $x \leq y$ implies $x = y$ (antisymmetry),
    \item $z \leq y$ and $y \leq x$ implies $z \leq x$ (transitivity). 
\end{enumerate}
We write $y < x$ if $y \leq x$ but $y \neq x$. An element $x \in\pos$
\emph{covers} $y \in\pos$ if $y < x$ and there is no $z \in \pos$ in-between, i.e., with
$y < z < x$ \cite{Rota:64}.

\mypar{Relation between DAGs and Posets} Every DAG $\dg = (\vrt,\edg)$ induces a unique partial order on $\vrt$, defined as $y < x$ if $y$ is a predecessor of $x$, i.e., if there is a path from $y$ to $x$. 

Conversely, for a given poset $P$ there are several DAGs that induce it, but two are special and unique. One is the {\em reachability} graph $\cl{\dg} = (\pos,\cl{\edg})$ with $\cl{\edg} = \{(y,x) \mid y < x \}$. This DAG is {\em transitively closed}, i.e., whenever there is a path from $y$ to $x$ there is also an edge $(y,x)$. We mark this property with an overline. The other unique DAG inducing $\pos$ is the {\em cover graph} $\dg = (\pos,\edg)$ with $\edg = \{(y,x) \mid x \text{ covers } y \}$. This graph is {\em transitively reduced}, i.e., the DAG with the fewest number of edges inducing $\pos$. It contains no edge $(y,x)$ if there is another path from $y$ to $x$ in $\dg$.

\begin{figure}
	\centering
	\subcaptionbox{\label{ex:GeneralDAG}}[0.3\linewidth]
	{\includegraphics[width=0.9\linewidth]{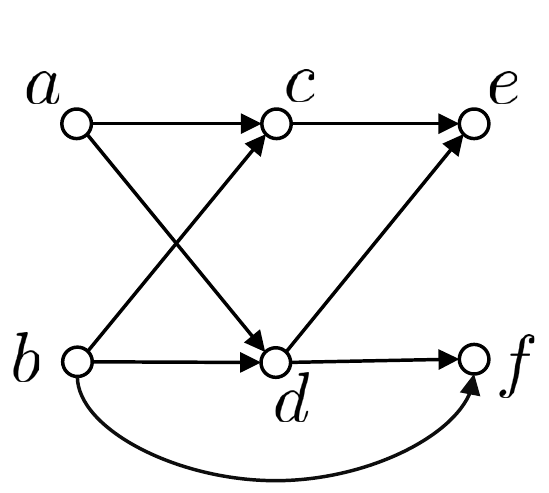}}
	\hfill
	\subcaptionbox{\label{ex:TransReducedDAG}}[0.3\linewidth]
	{\includegraphics[width=0.9\linewidth]{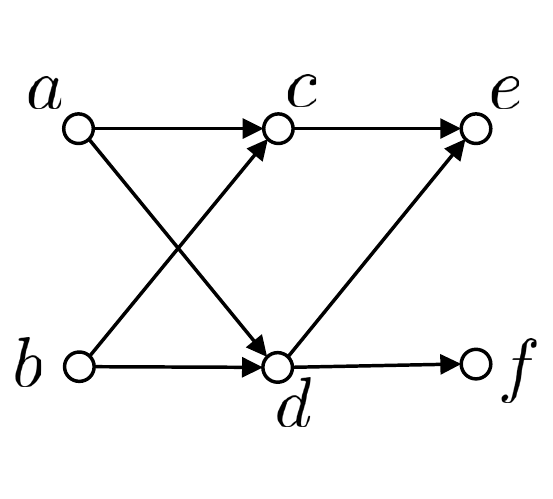}}
	\hfill
	\subcaptionbox{\label{ex:TransReducedDAG}}[0.3\linewidth]
	{\includegraphics[width=0.9\linewidth]{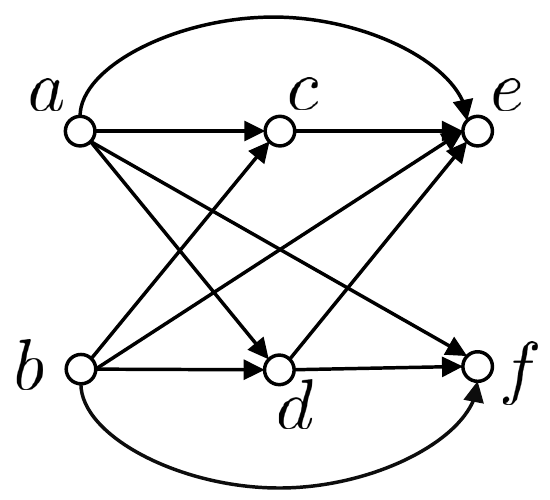}}
	\hspace*{\fill}%
	\caption{(a) A DAG $\dg$, (b) its transitive reduction, and (c) its transitive closure $\cl{\dg}$). All three induce the same poset for which (b) is the cover graph, and (c) the reachability graph.}
	\label{fig:ExampleDAG}
\end{figure}

\mypar{Example} In Fig.~\ref{fig:ExampleDAG} we show an example of a
DAG together with its transitive reduction and its transitive closure. All three DAGs induce the same poset.

In summary, every DAG uniquely defines a poset, whereas a poset can be represented by several DAGs. These however, have the same transitive reduction or transitive closure. Next we will build our signal model, which will require the computation of a transitive closure that also takes the weights into account.

\section{Signal Model and Weighted Transitive Closures}\label{sec:WeightedTransitiveClosure}%

We define the signal model that underlies our proposed Fourier analysis and the associated related SP concepts that we derive in the following section. The key aspect here is that the model, and thus the associated Fourier analysis, is not uniquely determined by the given weighted DAG $\dg$ but also requires a choice of weighted transitive closure that captures long-distance influences in $\dg$, i.e., for $y\leq x$ where $y$ is not a parent of $x$. We motivate the need for transitive closure and discuss relevant choices that depend on the meaning of the edge weights in $\dg$, leveraging the theory in \cite{Lehmann:1977a}.

\subsection{Basic signal model}\label{basic}

We assume a given weighted DAG $\dg = (\vrt,\edg,A)$ with induced partial order $\leq$ on $\vrt$, $|V| = n$. We consider signals on $\dg$ as column vectors of the form
$$
\coord{s} = (s_x)_{x\in\vrt}\in\R^n,
$$
where the order of the $s_x$ is determined by the chosen topological sort of $\vrt$.

We call each $x \in\vrt$ an event and say that an event $y$ is a cause of the event $x$ if $y \leq x$. Further, we assume that every event $y \in\vrt$ is associated with an unknown contribution (or input) $c_y \in \R$ to the DAG and that the (measured) signal value $s_x$ at $x\in\vrt$ is given by the weighted sum of the $c_y$ over all causes $y$:
\begin{equation}\label{eq:WeightedCausalSignal}
	s_x = \sum_{y \leq x} w_{x,y} c_y,\quad x\in\vrt.
\end{equation}
By (slight) abuse of notation we also say that $c_y$, for $y\leq x$, is a cause of $s_x$.

Intuitively, the weights in~\eqref{eq:WeightedCausalSignal} determine the influence of the causes of $x$ on $x$. Collecting the $w_{x,y}$ in a matrix yields the equivalent form
\begin{equation}\label{eq:matform} 
\coord{s} = W\coord{c},
\end{equation}
where $W$ is lower triangular and its nonzero entries correspond to edges in the reachability graph associated with $\dg$ (e.g., Fig.~\ref{fig:ExampleDAG}(c)).

\mypar{Example and motivation} As a simple, high level example (that we also used in the follow-up work \cite{misiakos2023learning}), consider a river network, which is a DAG $\dg$ since water only flows downstream. The nodes $x$ correspond to a set of fixed locations (e.g., cities). We assume that each node $y$ inserts an unknown amount $c_y$ of pollution, and that $s_x$ is the pollution measured at $x$, accumulated from all predecessors. The weight $a_{x,y}$ in $\dg$ could quantify what fraction of pollution at $y$ reaches a direct successor $x$. 

The measured pollution $s_x$ is determined by the $c_y$ of all causes $y\leq x$ of $x$ (not just the parents of $x$) and $w_{x,y}$ in \eqref{eq:WeightedCausalSignal} would capture their relative contribution. As we will explain below, $W$ is obtained from $A$ through a {\em weighted transitive closure}. Its exact form will depend on the meaning of the edge weights $a_{x,y}$.

\mypar{On the use of the term ``cause''} We use the term cause since we believe it helps with understanding our model. However, as already mentioned in the introduction, we want to stress again that \eqref{eq:WeightedCausalSignal} does not imply causality but could just express a linear relation, excluding hidden, confounding variables. Thus, strictly speaking, the term ``cause'' for the $c_y$ is not correct in this case. We still use it to emphasize that if \eqref{eq:WeightedCausalSignal} is a causal relationship, it does relate signal values and causes, and since it helps with understanding the different forms of transitive closure discussed next. In any event, our Fourier analysis and entire framework is applicable to any signal on any DAG.

\subsection{Weighted Transitive Closure}\label{sec:wtc}

The DAG $\dg$ and its edge weights captures how each node $x$ is influenced by its parents. But, as we also saw in the river network example, if $z$ is a parent of $x$ and $y$ is a parent of $z$, then, by transitivity, also $y$ will influence $x$. The associated weight $w_{x,y}$ will depend on the meaning of the edge weights in $\dg$. We build on the theory in \cite{Lehmann:1977a}.

\begin{figure}\centering
\includegraphics[width=0.336\linewidth]{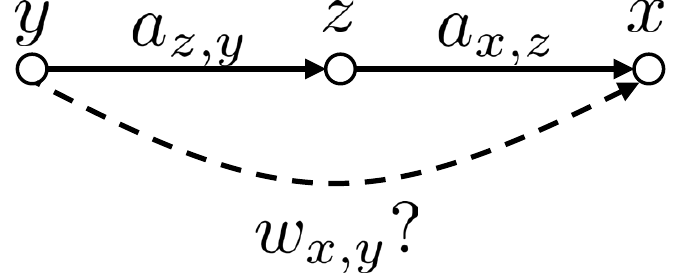}
\caption{The weighted transitive closure problem for a simple DAG with three nodes and two edges.\label{extc}}
\end{figure}

As a simple example, consider the transitive closure of the very small DAG in Fig.~\ref{extc}, i.e., the computation of $w_{x,y}$. In the river network example, fractions would multiply, i.e., $w_{x,y} = a_{z,y}a_{x,z}$. If the weights denoted distance, $w_{x,y} = a_{z,y} + a_{x,z}$, if they denoted throughput, $w_{x,y} = \min(a_{z,y},a_{x,z})$, and so on.

Next, we consider these and other choices that can be used to define $w_{x,y}$ for all $y < x$, given $\dg$. This is equivalent to computing a weighted transitive closure $\cl{\dg} = (\vrt, \cl{\edg}, \cl{A})$ of $\dg = (\vrt, \edg, A)$ and setting $w_{x,y} = \cl{a}_{x,y}$ for $y < x$. $(\vrt, \cl{\edg})$ is the reachability graph defining the same partial order as $(\vrt,\edg)$. Then we will define $w_{x,x}$ to obtain the entire matrix $W$ in \eqref{eq:matform}.


\mypar{Boolean weights: Standard transitive closure} If $A$ is the adjacency matrix, i.e., binary with all nonzero weights $=1$, one obvious choice is to give all edges in the transitive closure the weight $1$ as well, i.e., set $\cl{A}$ as the adjacency matrix of $\cl{\dg}$.

\mypar{Pollution} As in the example of the last section, the weights in $A$ could encode what fraction of a potential pollutant inserted at a node $y$ arrives at a direct successor $x$. Thus the weights are in $[0,1]$ and, for each node, the sum of the weights of outgoing edges should be $\leq 1$. The transitive closure $\cl{A}$ would then contain the same information, but now for each pair $(y,x)$ of nodes connected by a path. The fractions multiply along paths and have to be summed over all paths to obtain the result.

In this case there is a known formula~\cite{Lehmann:1977a}, which uses that $A^n = 0$:\footnote{The equation follows from the telescoping sum $\cl{A}(\one_n-A) = A - A^n = A$.}
\begin{equation}\label{eq:niceform}
\cl{A} = A + A^2 + \dots + A^{n-1} = (\one_n - A)^{-1} - \one_n.
\end{equation}

\mypar{Reliability/influence} The weights in $A$ could encode reliability or influence factors in $[0,1]$, where $a_{y,x} = 1$ means 100\% reliability of the edge or influence of $y$ on $x$ and $a_{y,x} = 0$ means none. Along paths, these influences multiply and $\cl{a}_{y,x}$ could encode the most reliable/influential path from $y$ to $x$.

\mypar{Shortest path} The weights in $A$ could encode distances in $\R^+$ between nodes. In this case, the weights $\cl{a}_{y,x}$ in $\cl{A}$ could be defined as the shortest path from $y$ to $x$ in $\dg$. Since one would assume causes $y$ that are farther from $x$ in $\dg$ to have less influence, one could consider derivatives of a path length $\ell$ such as $1/\ell$ or $e^{-\ell}$. The latter choice effectively converts distances to influences in the sense discussed right above.

\mypar{Maximal capacity} The weights in $A$ could encode capacity or throughput $\in\R^+$ of edges. The capacity of a path is determined by the minimal capacity among its edges and $\cl{a}_{y,x}$ could encode the maximal capacity path between $y$ and $x$. 

\begin{table*}[t!]
	\centering
	\begin{tabular}{lllllll}
		\toprule
		{$S$} & {$u \oplus v$} & {$u \odot v$} & {$0_S$} & {$1_S$}
		& {Meaning of edge weight $\cl{a}_{x,y}$ in closure}  \\ 
		\midrule
		$\{ 0,1\}$ & $u \text{ or } v$ & $u \text{ and } v$ & $0$ & $1$ &
		$x$ is reachable from $y$, i.e., $y\leq x$
		\\
		$[0,1]$ & $u + v$ & $u \cdot v$
		& $0$ & $1$ & Fraction of pollution from $y$ reaching $x$ \\
		$[0,1]$ & $\max(u,v)$ & $u \cdot v$
		& $0$ & $1$ & Strongest influence/most reliable path from $y$ to
		$x$\\    
		$\mathbb{R}^+ \cup \{\infty\}$ & $\min(u,v)$ & $u + v$ 
		& $\infty$ & $0$ & Shortest path length from $y$ to $x$ \\
		$\mathbb{R}^+ \cup \{\infty\}$ & $\max(u,v)$ & $\min(u,v)$ 
		& $0$ & $\infty$ & Largest capacity path from $y$ to $x$\\  
		\bottomrule
	\end{tabular}
	\caption{Examples of choices for semiring operations $\oplus,\otimes$ when operating on edge weights in the algorithm of Fig.~\ref{algo:ModifiedFloydWarshall} and the associated meaning of the edge weights. For the pollution interpretation in the second row, the weights of outgoing edges have to sum to $\leq 1$. The table is adapted from~\cite{Abdali.Saunders:1985a}.}
	\label{tab:ClosedSemirings}
\end{table*}

\mypar{Computation} Various algorithms are available to compute transitive closures and their associated weights. The special cases that we just presented can be solved with one generic algorithm, instantiated in different ways~\cite{Abdali.Saunders:1985a}. We show it in its simplest form in Fig.~\ref{algo:ModifiedFloydWarshall}; an optimized version can be found in~\cite{Abdali.Saunders:1985a}. The algorithm is initialized with the weight matrix $A$ on which it performs $n^3$ iterations for a total runtime of $O(n^3)$. It is generic in the choice of addition $\oplus$ and multiplication $\otimes$ used, which must satisfy a semiring property\footnote{See~\cite[Definition 2.1]{Abdali.Saunders:1985a}. In particular, $\oplus$ is commutative, $\oplus$ and $\otimes$ are associative, have an identity element, and satisfy the distributivity law.}.

\begin{figure}\footnotesize
	\centering
	\begin{algorithmic}
		\Function{\textbf{WeightedTransitiveClosure}}{$A$}
		\State $H^{(0)} \leftarrow A$ 
		\For{$k = 1,\dots,n$}
		\For{$i = 1,\dots, n$}
		\For{$j = 1,\dots, n$}
		\State $h^{(k)}_{i,j} \leftarrow h^{(k-1)}_{i,j} \oplus
		( h^{(k-1)}_{i,k} \odot h^{(k-1)}_{k,j})$
		\EndFor
		\EndFor
		\EndFor
		\State \textbf{return} $\cl{A} = H^{(n)}$
		\EndFunction 
	\end{algorithmic}    
	\caption{Generic algorithm to compute various forms of weighted transitive closure of $A$ in $O(n^3)$~\cite{Abdali.Saunders:1985a}. The genericity is in the choice of addition $\oplus$ and multiplication $\otimes$, which need to satisfy a semiring property. Possible choices and the associated results are shown in Table~\ref{tab:ClosedSemirings}.}
	\label{algo:ModifiedFloydWarshall}
\end{figure}

\begin{figure}\centering
	\subcaptionbox{Product: Combing consecutive weights along a path.\label{ex:TransReducedDAG}}[0.48\linewidth]
	{\includegraphics[width=0.7\linewidth]{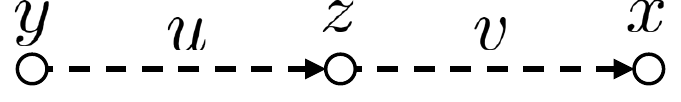}}
	\hfill
	\subcaptionbox{Sum: Combining weights over alternative paths.\label{ex:TransReducedDAG}}[0.48\linewidth]
	{\includegraphics[width=0.7\linewidth]{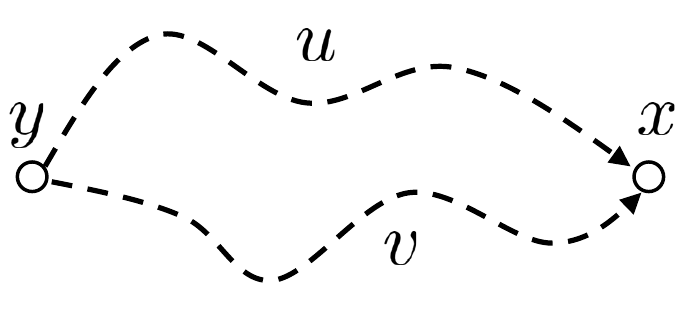}} 
	\hfill   
	\caption{The purpose of product and sum in Table~\ref{tab:ClosedSemirings}. If these operations satisfy the semiring property, the algorithm in Fig.~\ref{algo:ModifiedFloydWarshall} works.}
	\label{fig:prodsum}
\end{figure}

Table~\ref{tab:ClosedSemirings} shows several choices of semirings $S$ and the associated result of the algorithm. Fig.~\ref{fig:prodsum} provides intuition: $\otimes$ determines how consecutive weights are combined (e.g., product for reliability, sum for path length), and $\oplus$ how weights of alternative paths are combined (e.g., sum for pollution, min for shortest path length). If these two operations satisfy the semiring property, then the algorithm in Fig.~\ref{algo:ModifiedFloydWarshall} works.

Note that with $\oplus$ and $\otimes$ also the definition of $0$ (identity element for addition) and $1$ (identity element for multiplication) changes as shown in the table. E.g., for shortest path, $0_S = \infty$ since $u\oplus \infty = \min(u,\infty) = u = \infty\oplus u$. Thus, in the algorithm, zeros in $A$ have to be replaced with $\infty$ upon initialization.

For shortest path, the algorithm is equivalent to the classical Floyd-Warshall algorithm~\cite{Floyd:1962a}. 

Other choices of weighted transitive closure may require other algorithms. E.g., overall capacity between two nodes requires max-flow algorithms \cite{Goldberg:1988}. The related structural equation models (discussed later in Section~\ref{sec:dis}) use the pollution model but without constraints on the weights.

\begin{figure}
	\centering
	\subcaptionbox{Boolean \label{ex:ExampleDAGBoolean}}[0.35\linewidth]
	{\includegraphics[width=0.8\linewidth]{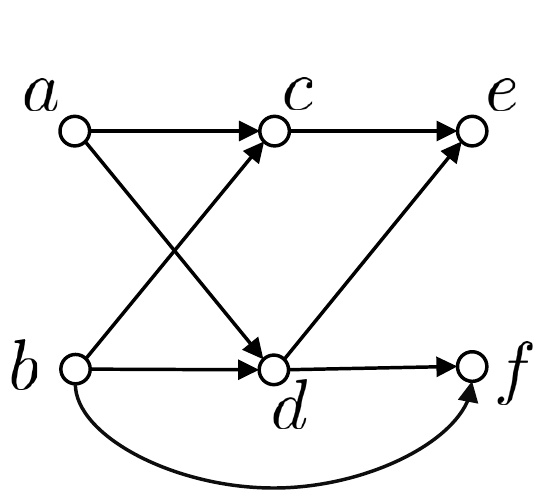}}
	\subcaptionbox{Closed\label{ex:ExampleDAGBooleanClosed}}[0.35\linewidth]
	{\includegraphics[width=0.8\linewidth]{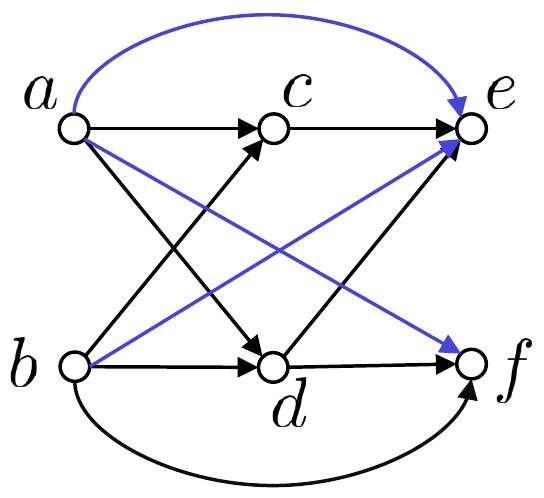}}
	\subcaptionbox{Pollution \label{ex:ExampleDAGPollution}}[0.35\linewidth]
	{\includegraphics[width=0.8\linewidth]{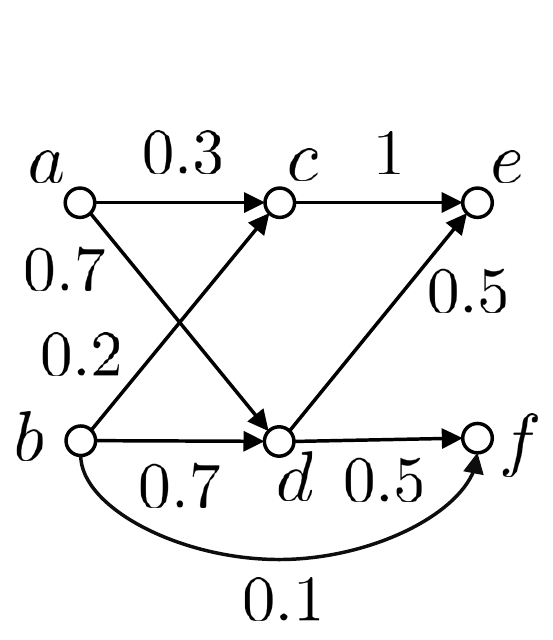}}
	\subcaptionbox{Closed\label{ex:ExampleDAGPollutionClosed}}[0.35\linewidth]
	{\includegraphics[width=0.8\linewidth]{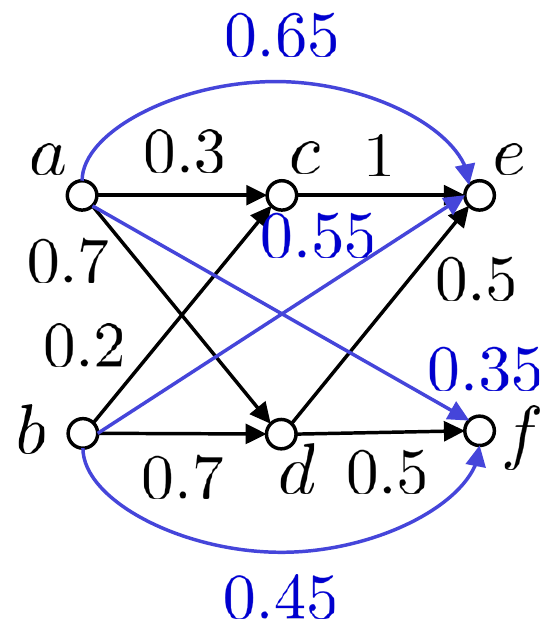}}
	\subcaptionbox{Shortest paths\label{ex:ExampleDAGShortestPaths}}[0.35\linewidth]
	{\includegraphics[width=0.8\linewidth]{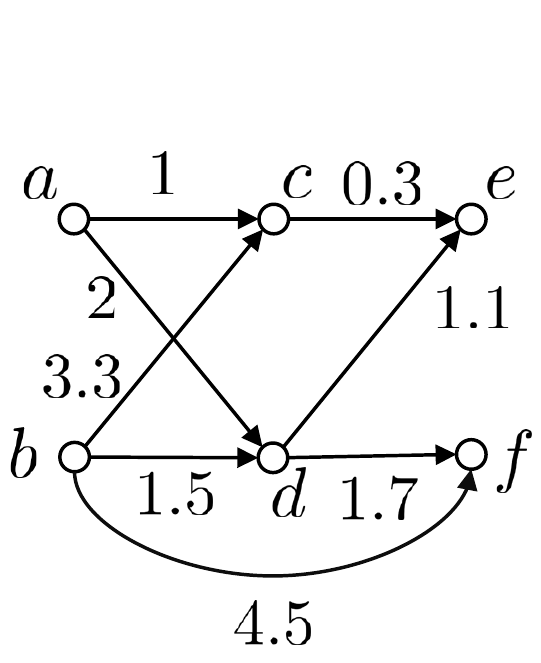}}
	\subcaptionbox{Closed\label{ex:ExampleDAGShortestPathsClosed}}[0.35\linewidth]
	{\includegraphics[width=0.8\linewidth]{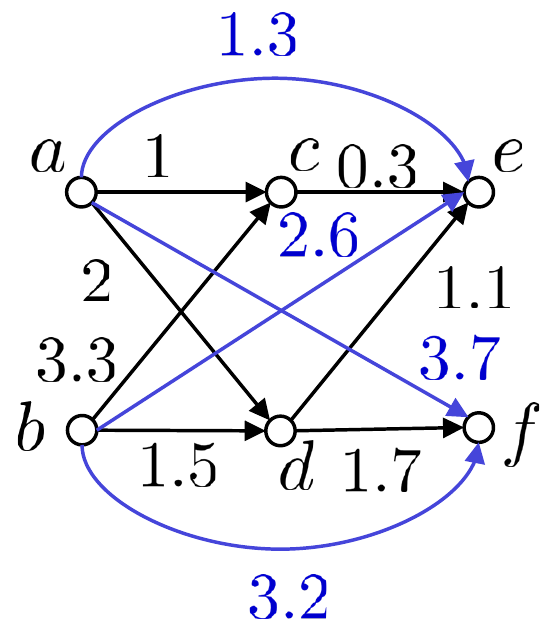}}
	\subcaptionbox{Influence \label{ex:ExampleDAGInfluence}}[0.35\linewidth]
	{\includegraphics[width=0.8\linewidth]{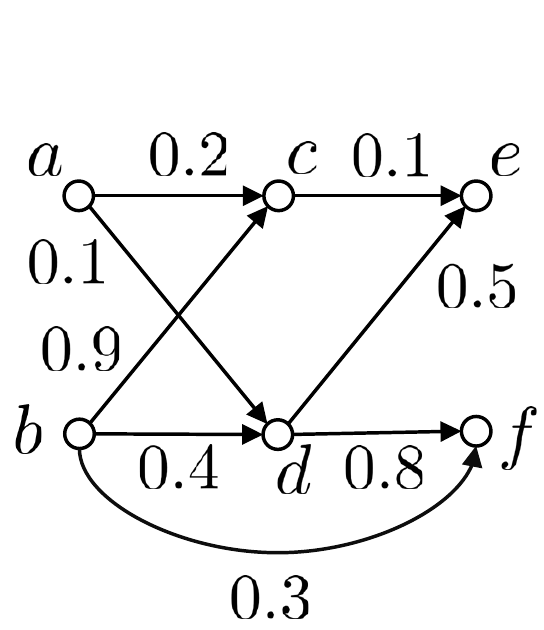}}
	\subcaptionbox{Closed\label{ex:ExampleDAGInfluenceClosed}}[0.35\linewidth]
	{\includegraphics[width=0.8\linewidth]{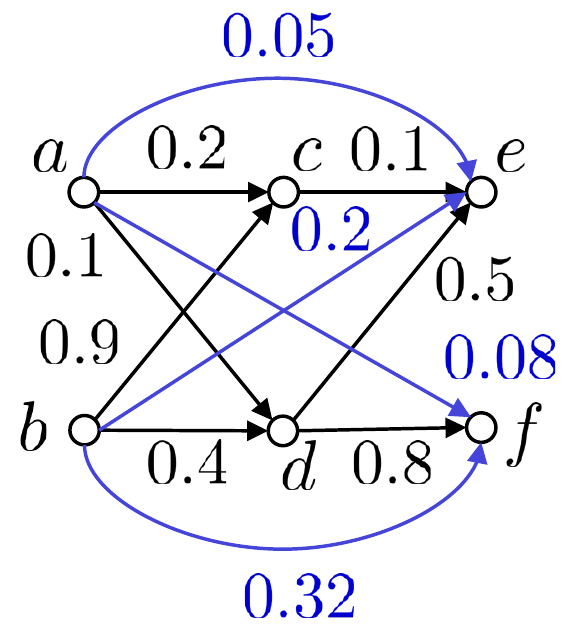}}
	\caption{Example DAG with different weights and meanings of weights (left) and their corresponding	transitive closures (right). Modified and added weights and edges are colored blue.}
	\label{fig:ExampleDAGWeights}
\end{figure}

\mypar{Examples} Fig.~\ref{fig:ExampleDAGWeights} shows a few examples of transitive closures, using the DAG from Fig.~\ref{fig:ExampleDAG}(a) as starting point. Note that the transitive closure $\cl{A}$ may overwrite weights in $A$. E.g., the bottom edge in Fig.~\ref{ex:ExampleDAGShortestPaths} has weight $4.5$, but after closure, the shortest path from $b$ to $f$ has length $3.2 = 1.5 + 1.7$.

\subsection{Reflexive closure}

Finally, we need to define $w_{x,x}$ in \eqref{eq:WeightedCausalSignal}, which accounts for reflexivity in the partial order, i.e., the fact that $c_x$ is a cause of $s_x$. Our model requires $w_{x,x}\neq 0$, since later we want the triangular matrix $W$ in \eqref{eq:matform} to be invertible. In this paper we focus on the choice $w_{x,x}=1$. For the special case of pollution, we obtain a closed form using \eqref{eq:niceform}:
\begin{equation}\label{eq:nicew}
W = \cl{A} + \one_n = (\one_n - A)^{-1}.
\end{equation}
In other cases, $W$ takes a different forms as computed by the algorithm in Fig.~{algo:ModifiedFloydWarshall}. 

In three of the fives cases in Table~\ref{tab:ClosedSemirings} the choice of $1$ coincides with $1_S$ (fraction of pollution or influence of a node on itself is $=1$), but poses a problem for the others. For shortest paths this suggests, for example, and as we also do later, converting path length $d$ to exponential decay $e^{-d}$, which makes it a particular influence model, effectively converting the operations $(\min,+)$ to $(\max,\cdot)$. For capacity one could choose $e^{-1/c}$ for capacity $c$.

\mypar{Summary} Both $A$ and $\cl{A}$ are lower triangular matrices with zeros on the diagonal and thus as the only eigenvalue. $W = \one_n+\cl{A}$ is lower triangular with ones on the diagonal and thus of full rank, i.e., its columns form a basis, which, in fact, will become our proposed Fourier basis as explained in the next section.

\section{Causal Fourier Analysis on DAGs}\label{sec:WeightedCausalSP}

Given a weighted DAG $\dg = (\vrt, \edg, A)$ with associated partial order $\leq$, we assume we have decided on a suitable transitive/reflective closure $W$ of $A$ as explained in Section~\ref{sec:WeightedTransitiveClosure}. We restate our signal model, which assumes that a signal on $\dg$ is a linear combination of unknown causes
associated with the nodes:
\begin{equation}\label{eq:modelagain}
	s_x = \sum_{y \leq x} w_{x,y} c_y\quad\text{or}\quad\coord{s} = W\coord{c}.
\end{equation}
In this section we will build on this equation to develop a linear SP framework for signals on DAGs. In short, we will argue that $\coord{c}$ can be interpreted as a form of spectrum of $\coord{s}$, with $W^{-1}$ as associated Fourier transform. We do so following the general theory in \cite{Pueschel:08a,Pueschel:06c}: we define a suitable notion of shift and convolution for which the columns of $W$ form a joint eigenbasis. Our derivations leverage the classical theory of Moebius inversion from combinatorics \cite{Rota:64}, which is concerned with equations on posets of the form in \eqref{eq:modelagain}. We start by inverting \eqref{eq:modelagain}.

\subsection{Calculating Causes: Moebius Inversion}

In the case of the standard Boolean transitive closure, i.e., trivial nonzero weights $w_{x,y} = 1$, the calculation of $\coord{c}$ from $\coord{s}$ is provided by the classical Moebius inversion from \cite{Rota:64}. The extension to arbitrary transitive closures and weights needed here is straightforward and provides a formula for $W^{-1}$.

\begin{theorem}\label{th:wmi}
\begin{equation}
    \label{eq:WeightedMoebiusInversion}
    s_x = \sum_{y \leq x} w_{x,y} c_y,
    \text{ if and only if }
    c_y = \sum_{x \leq y} \mu_w(x,y) s_x.
\end{equation}
Here $\mu_w$ is the weighted Moebius function, recursively defined as
\begin{align*}
  \label{eq:WeightedMoebiusFunction}
  \mu_w(x,x) &= 1, &\text{ for } x \in \vrt, \\
  \mu_w(x,y) &= - \sum_{x \leq z < y} w_{y,z} \mu_w(x,z),
                   &\text{ for } x \not= y. 
\end{align*}
\end{theorem}

We provide a proof in the appendix. $W^{-1}$ is lower triangular with ones on the diagonal since the same holds for $W$.


\subsection{From Shift to Fourier Transform}

The following definition of the shift operation is a key contribution of this paper. At first glance it is non-obvious but is the one that generalizes our prior SP framework on lattices~\cite{Pueschel.Seifert.Wendler:2020a} to arbitrary posets and weighted DAGs. As we will see, it can be viewed as a form of causal delay, takes an intuitive form in the case of Boolean weights, makes the columns of $W$ the associated Fourier basis, and thus the causes become the spectrum. Once the shifts are defined, the derivation of the remaining basic SP concepts is straightforward~\cite{Pueschel:08a}.

\mypar{Causal shifts} We first provide the formal shift definition and then interpret it. For every $q \in\vrt$ we define a linear shift operator, given by a matrix $T_q$, on $\coord{s}$:
\begin{equation}
    \label{eq:WeightedShiftFourierSpace}
    (T_q \coord{s})_x
    = \sum_{y \leq x \text{ and } y \leq q} w_{x,y} c_y
    \text{ for all } x \in\vrt.
\end{equation}
In words, comparing to \eqref{eq:modelagain}, the result is the signal with all causes removed, which are not common causes of $q$ and $x$. 

But to obtain a proper representation of this linear mapping we need to express the right-hand side of \eqref{eq:WeightedShiftFourierSpace} as linear combination of signal values, not causes. We do this by replacing $c_y$ in \eqref{eq:WeightedShiftFourierSpace}
using the inversion formula in \eqref{eq:WeightedMoebiusInversion}:
\begin{equation}
	\label{eq:WeightedShiftPosetSpace}
	(T_q \coord{s})_x
	= \sum_{y \leq x \text{ and } y \leq q} w_{x,y} \sum_{z \leq y} 
	\mu_w(z,y) s_z. 
\end{equation}
Inspecting \eqref{eq:WeightedShiftPosetSpace} shows that $(T_q \coord{s})_x$ is a linear combination of signal values $s_z$ with $z\leq x$ and $z\leq q$ (i.e., ``earlier'' in the DAG order), as visualized in Fig.~\ref{ex:ExampleDAGBooleanClosed}. Thus we consider it as a form of ``causal delay.'' 

\begin{figure}\centering
\subcaptionbox{Generic case\label{ex:ExampleDAGBooleanClosed}}[0.24\textwidth]
{\includegraphics[width=0.2\textwidth]{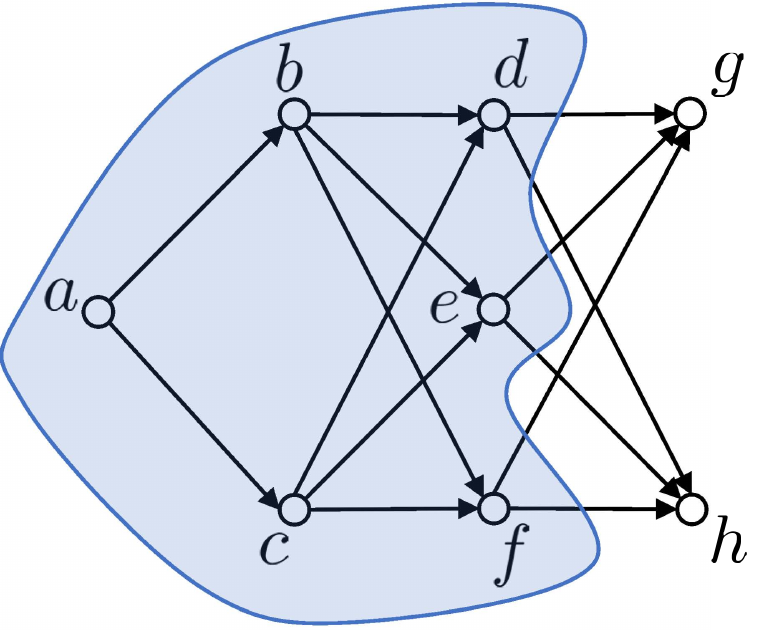}}
\subcaptionbox{Unique largest lower bound $e = g\meet h$.\label{lattshift}}[0.24\textwidth]
{\includegraphics[width=0.2\textwidth]{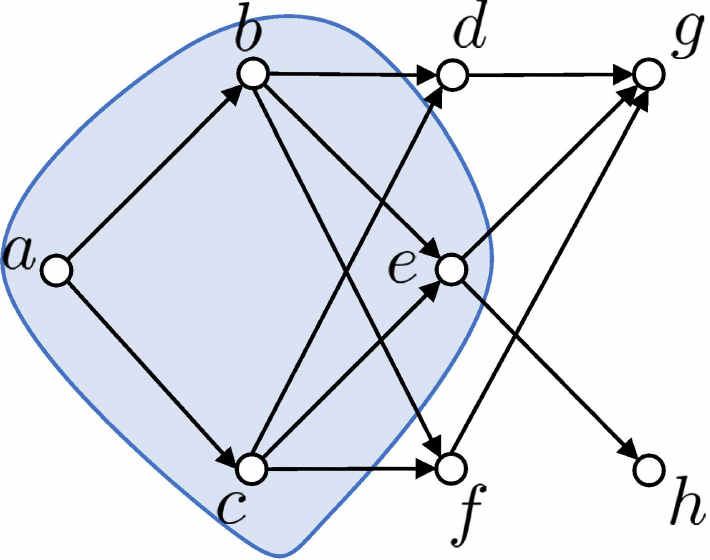}}\\
\subcaptionbox{Unique largest lower bound $e = g\meet h$ and Boolean weights.\label{lattshiftnw}}[0.24\textwidth]
{\includegraphics[width=0.2\textwidth]{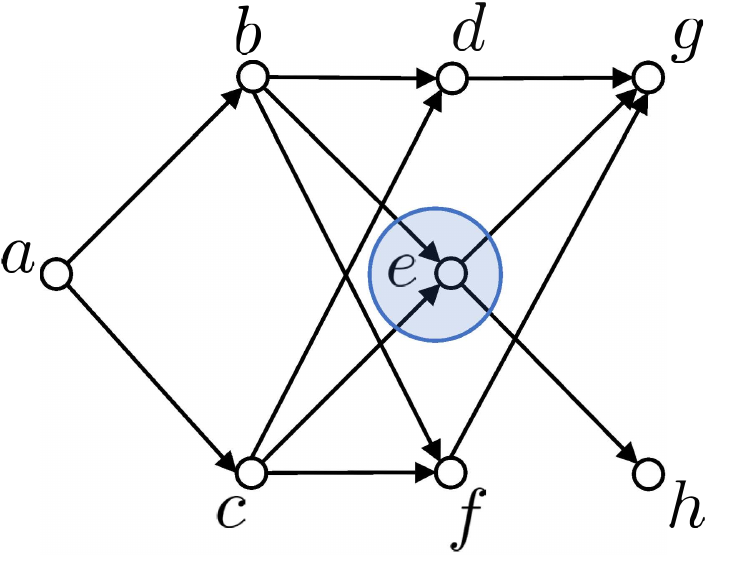}}
\caption{$\coord{s}$ shifted with $h$ at node $g$ (i.e., $(T_h \coord{s})_g$). (a) Generic case: linear combination of common predecessors of $g$ and $h$ such that common causes are removed. (b) Special case in which $g$ and $h$ have a unique largest lower bound in the partial order: linear combination of predecessors of $e$. (c) As in (b) but in addition with Boolean edge weights: $(T_h \coord{s})_g = s_e$.}
\label{fig:ShiftCauses}
\end{figure}

We consider special cases to motivate the definition. Assume that $x$ and $q$ have a unique greatest lower bound in $\dg$ denoted with $x\meet q$, i.e.: for all $y \neq x \meet q$ with $y\leq x\text{ and }y\leq q$ we have $x\meet q < y$ (which is the case if the poset is even a lattice \cite{Rota:64}). In this case, \eqref{eq:WeightedShiftPosetSpace} simplifies to
\begin{equation}\label{splatt}
	(T_q \coord{s})_x = \sum_{y \leq x\meet q} w_{x,y} \sum_{z \leq y} 
	\mu_w(z,y) s_z, 
\end{equation}
i.e., it is linear combination of signal values of nodes $\leq x\meet q$. This situation is visualized in Fig.~\ref{lattshift} with $e = g\meet h$.

Assume in addition that $W$ is the Boolean transitive closure of an unweighted DAG given by $A$, which is the situation in the prior work~\cite{Pueschel.Seifert.Wendler:2020a}. Then, now by specializing \eqref{eq:WeightedShiftFourierSpace},
\begin{equation}
	 (T_q \coord{s})_x = \sum_{y \leq x\meet q} c_y = s_{x\meet q},
\end{equation}
i.e., the causal delay takes its most beautiful form (in Fig.~\ref{lattshiftnw} the result is $s_e$) and can be conceptually compared to the classical shift of a discrete-time signal by $k$, which maps $s_n$ to $s_{n-k}$.

%

Equation~\eqref{eq:WeightedShiftFourierSpace} shows that all shift matrices $T_q$, $q\in\vrt$, commute since they only affect the range of summation, and that they are idempotent, i.e., $T_q\cdot T_q = T_q$. Further, \eqref{eq:WeightedShiftPosetSpace} shows that signal values are shifted to linear combination of predecessors; thus the $T_q$ are lower triangular and in general not invertible.

\mypar{Filters and convolution} The shifts generate the algebra (ring and vector space) of filters~\cite{Pueschel:08a}. Since the shifts commute, a filter is a polynomial in the shifts. Since $T_q^2 = T_q$, the most general filter, represented as matrix, is $\sum_{q \in\vrt} h_q T_q$ for $h_q\in\R$. Thus, for $\coord{h} = (h_q)_{q\in\vrt}\in\R^n$, the associated convolution takes the form
\begin{equation}\label{eq:Filter}
    \coord{h} * \coord{s}
    = \Big(\sum_{q \in\vrt} h_q T_q \Big) \coord{s}.
\end{equation}
As polynomials in the shifts, filters are shift-invariant, i.e., $\coord{h} * T_q \coord{s} = T_q(\coord{h} * \coord{s})$ for all $q\in\vrt$.

\mypar{Fourier basis and  transform} The Fourier basis consists of the joint
eigenvectors of all $T_q$ and thus all filters. Its derivation is simple due the definition of $T_q$. Let $\coord{s} = W\coord{c}$. Equation~\eqref{eq:WeightedShiftFourierSpace} shows that shifting $\coord{s}$ by $q$ performs a pointwise multiplication on $\coord{c}$. Formally, we can write \eqref{eq:WeightedShiftFourierSpace} as 
\begin{equation}\label{eq:matdiag}
T_q\coord{s} = W D_q\coord{c},\quad D_q = \diag_{y\in\vrt}(\chr{y\leq q}),
\end{equation}
where $\chr{y\leq q}$ is the indicator function
\begin{equation}\label{eq:indfunc}
    \chr{y \leq q} =
    \chr{y \leq q}(y,q) =
    \begin{cases}
        1 & \text{ if } y \leq q, \\
        0 & \text{ else.}
    \end{cases}
\end{equation}
In words, $D_q$ removes the causes of any signals that are not also causes of $q$.

Replacing $\coord{s} = W\coord{c}$ in \eqref{eq:matdiag} shows that for all $\coord{c}\in\R^n$ we have $T_qW\coord{c} = WD_q\coord{c}$, and thus
$$
T_qW = WD_q.
$$
Since $W$ has full rank, the columns are the desired Fourier basis:
\begin{theorem}[Fourier basis]
The columns of $W$ from a simultaneous eigenbasis of all shifts and filters, i.e., the Fourier basis vectors are
\begin{equation}\label{eq:FourierBasis}
	\coord{f}^y	= (w_{x,y})_{x \in\vrt},\quad y\in\vrt.
\end{equation}
\end{theorem}
The associated Fourier transform is obtained by inversion:
\begin{theorem}[Fourier transform]
The Fourier transform associated with the above Fourier basis is given by
$$
\ft{\coord{s}} = \coord{c} = W^{-1}\coord{s} = \Fourier_\dg\coord{s},
$$
with Fourier transform matrix (Theorem~\ref{th:wmi})
$$
\Fourier_\dg = W^{-1} = [\mu_w(x,y) \chr{x \leq y}]_{y,x \in\vrt}.
$$
\end{theorem}

\mypar{Frequency response and convolution theorem} Equation~\eqref{eq:matdiag} shows that the frequency response of a shift $T_q$ is the diagonal of $D_q$, i.e., $(\chr{y\leq q})_{y\in\vrt}$ by \eqref{eq:matdiag}. Thus, for a general filter $\coord{h}$ corresponding to the matrix $H = \sum_{q \in\vrt} h_q T_q$, it is the diagonal of $\sum_{q \in\vrt} h_q D_q$. 

Denoting the frequency response of $\coord{h}$ as $\coord{h}'$ we obtain
\begin{equation}\label{eq:FrequencyResponse}
    \fr{h}_y = \sum_{q \geq y} h_q,\quad y\in\vrt,
\end{equation}
and the associated convolution theorem:
\begin{equation}\label{eq:ConvolutionTheorem}
	\ft{\coord{h} * \coord{s}}
	= \fr{\coord{h}} \odot \ft{\coord{s}},
\end{equation}
where $\odot$ denotes pointwise multiplication. In words, convolution (filtering) in the signal domain is equivalent to pointwise multiplication (by the frequency response of the filter) in the frequency domain. A few things are worth noting.

The Fourier transform and frequency response are computed differently, which is also the case in graph signal processing and generally due to the different roles of signal and filter space \cite{Pueschel:08a}. 

The frequency response is computed only based on the partial order defined by the DAG, i.e., independent of the weights and the chosen transitive closure. 

Computing $\coord{h}'$ from $\coord{h}$ is a linear transform with an lower triangular transform matrix with ones on the diagonal. As such it is invertible, which means every frequency response can be achieved with a suitable filter. In particular, the trivial filter with $\fr{h}_y = 1$, $y\in\vrt$, i.e., $H = \one_n$ is a suitable linear combination of the $T_q$. The inversion of the frequency response can be done with the dual version of Theorem~\ref{th:wmi}, i.e., in which $\leq$ is replaced with $\geq$.

\mypar{Fast algorithms} Explicitly computing the Fourier transform matrix $\Fourier_\dg = W^{-1}$ by inverting $W$ requires $O(n^3)$ operations. In practice, since the matrix and its inverse are lower triangular, the spectrum $\ft{\coord{s}}$ of a signal $\coord{s}$ can be computed without explicitly constructing $\Fourier_\dg$ using a triangular solve of
\begin{equation}\label{eq:TriangularSolve}
	W \ft{\coord{s}} = \coord{s},
\end{equation}
using $O(n^2)$ operations.

In the case of unweighted (i.e., Boolean weights) DAGs the Fourier transform and its inverse can be computed using $O(nk)$ time and memory, where $k$ is the width of the DAG, i.e., the longest antichain or maximal number of mutually non-comparable elements of the DAG~\cite{Pegolotti.Seifert.Pueschel:2022a}. For small $k$ this enables the computation for DAGs up to millions of nodes.

\mypar{Total variation and frequency ordering} We complete our framework by a suitable definition of frequency ordering. Here we follow the high-level idea used for graphs by~\cite{Sandryhaila:13}, which relates frequency ordering to the shift via total variation (TV). Here, however, we have multiple shifts and thus we consider TV separately for each shift, as common for images (with two shifts: horizontal and vertical translation) and as was done previously in \cite{Pueschel.Seifert.Wendler:2020a} for meet/join lattices, which thus we generalize here.
 
\begin{definition}\label{def:PosetTotalVariation}%
    Let $\coord{s}$ be a signal on $\dg$. We define the variation w.r.t.~a shift by $q$ as 
    $\TV_q(\coord{s}) = \norm{\coord{s} - T_q \coord{s}}_2$. The total
    variation of $\coord{s}$ is then the vector
    \begin{equation}\label{eq:TotalVariation}
        \TV(\coord{s}) = (\TV_q(\coord{s}))_{q \in\vrt}
    \end{equation}
    and the sum total variation is the number
    \begin{equation}\label{eq:SumTotalVariation}
        \STV(\coord{s}) = \sum_{q \in\vrt} \TV_q(\coord{s}).
    \end{equation}
\end{definition}
We next show that the Fourier basis, and thus the spectrum, is partially ordered in a way isomorphic to the partial order induced by $\dg$, with low frequencies corresponding to the early nodes in the DAG, i.e., the smallest ones in the induced partial order.
\begin{theorem}\label{thm:PropsTotalVariation}%
    We normalize $\coord{f}^y$ to
    $\norm{\coord{f}^y}_2 = 1$. Then 
    \begin{equation}\label{eq:TotalVariationOfNormalizedEV}
        \TV(\coord{f}^y) = (\chr{y \not\leq q})_{q \in\vrt}
    \end{equation}
    and thus
    \begin{equation}
        \label{eq:SumTotalVariationOfNormalizedEV}
        \STV(\coord{f}^y) = \norm{\TV(\coord{f}^y)}_1 = |\{ q\in\vrt \; | \; y \not\leq q\}|.
    \end{equation}
    The poset of total variations
    $T = \{ \TV(\coord{f}^y) \; | \; y \in \vrt \}$ w.r.t componentwise
    comparison is isomorphic to the (unweighted) poset induced by $\dg$, i.e., 
    $x \leq y$ if and only if
    $\TV(\coord{f}^x) \leq \TV(\coord{f}^y)$. STV provides a topological sort of the frequencies.
\end{theorem}

We provide a proof in the appendix. Note that the frequency ordering only depends on the partial order induced by the DAG and not by the chosen weighted transitive closure. Also, the frequency ordering is independent of the choice of the two norms occurring in Theorem~\ref{thm:PropsTotalVariation}.

Next we provide a numerical example for the concepts introduced and then we conclude this section by a discussion of salient aspects of our framework.


\subsection{Small Example}

For a small example we consider the DAG $\dg$ in Fig.~\ref{ex:ExampleDAGPollution}, which is given by
$$
A = 
\left[\footnotesize\ra{1.0}
\begin{array}{@{}rrrrrr@{}}
	0 & 0 & 0 & 0 & 0 & 0\\
	0 & 0 & 0 & 0 & 0 & 0 \\
	0.3 & 0.2 & 0 & 0 & 0 & 0\\
	0.7 & 0.7 & 0 & 0 & 0 & 0\\
	0 & 0 & 1 & 0.5 & 0 & 0 \\
	0 & 0.1 & 0 & 0.5 & 0 & 0
\end{array}
\right].
$$
Further, we assume the pollution model, i.e., the transitive closure of $A$ is given by Fig.~\ref{ex:ExampleDAGPollutionClosed}, which, including the reflexive closure, yields the matrix
$$
W = 
\left[\footnotesize\ra{1.0}
\begin{array}{@{}rrrrrr@{}}
	1 & 0 & 0 & 0 & 0 & 0\\
	0 & 1 & 0 & 0 & 0 & 0 \\
	0.3 & 0.2 & 1 & 0 & 0 & 0\\
	0.7 & 0.7 & 0 & 1 & 0 & 0\\
	0.65 & 0.55 & 1 & 0.5 & 1 & 0 \\
	0.35 & 0.45 & 0 & 0.5 & 0 & 1
\end{array}
\right].
$$

\begin{figure}\centering
\includegraphics[width=0.2\textwidth]{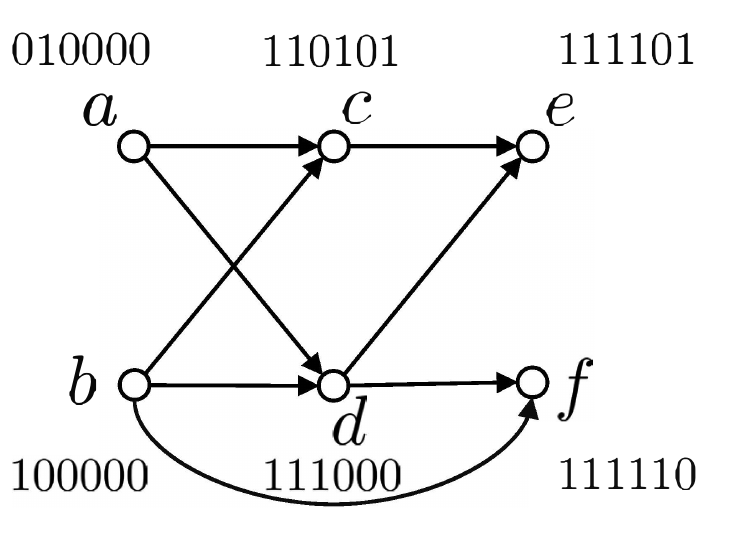}
\caption{Partial ordering of the spectrum of the DAG in Fig.~\ref{ex:ExampleDAGPollution}. $\TV(\coord{f}^y)$ is shown next to node $y$. \label{fig:TVex}}
\end{figure}

\mypar{Fourier basis and frequency ordering} The columns of $W$ constitute the Fourier basis $\{\coord{f}^a,\dots,\coord{f}^f\}$ of the DAG. The first two have the lowest frequency w.r.t.~the total variation in Definition~\ref{def:PosetTotalVariation}. For example $\TV(\coord{f}^a)= (\chr{a \not\leq q})_{q \in\vrt} = (0,1,0,0,0,0)$. All $\TV(\coord{f}^y)$ are shown in Fig.~\ref{fig:TVex}. Note, for example, that $\TV(\coord{f}^c)\leq \TV(\coord{f}^e)$ but $\TV(\coord{f}^c)\not\leq \TV(\coord{f}^f)$.

\mypar{Fourier transform} The inverse Fourier transform connecting signal and causes (spectrum) is given by $\coord{s} = \Fourier_\dg^{-1}\ft{\coord{s}} = W\coord{c}$. Thus the Fourier transform
becomes $\ft{\coord{s}} = \Fourier_\dg\coord{s} = W^{-1}\coord{s}$.
%
%
The Fourier transform matrix is given by $\Fourier_\dg = W^{-1}$:
\begin{equation}\label{eq:ExamplePosetFT}
\Fourier =
\left[\footnotesize\ra{1.0}
\begin{array}{@{}rrrrrr@{}}
    1 & 0 & 0 & 0 & 0 & 0 \\
    0 & 1 & 0 & 0 & 0 & 0 \\
    -0.3 & -0.2 & 1 & 0 & 0 & 0 \\
    -0.7 & -0.7 & 0 & 1 & 0 & 0 \\
    0 & 0 & -1 & -0.5 & 1 & 0 \\
    0 & -0.1 & 0 & -0.5 & 0 & 1
\end{array}
\right]. 
\end{equation}

\mypar{Shifts} As an example, the shift $T_e$ is given by the matrix
\begin{equation}
\label{eq:ExamplePosetShiftMatrix}
T_e =
\left[\footnotesize\ra{1.0}
\begin{array}{@{}rrrrrr@{}}
    1 & 0 & 0 & 0 & 0 & 0 \\
    0 & 1 & 0 & 0 & 0 & 0 \\        
    0 & 0 & 1 & 0 & 0 & 0 \\        
    0 & 0 & 0 & 1 & 0 & 0 \\        
    0 & 0 & 0 & 0 & 1 & 0 \\
    0 & 0.1 & 0 & 0.5 & 0 & 0  
\end{array}
\right]. 
\end{equation}
For example, 
\begin{equation}
    \label{eq:ExamplePosetShiftingValues}
    \begin{split}
        (T_e \coord{s})_f
        &= 0.1 s_b + 0.5 s_d
    \end{split}
\end{equation}
which is a linear combination  of signal values on common predecessors of $e$ and $f$ in $\dg$.

\mypar{Low-pass filter} In classical discrete-time SP, a basic
low-pass filter is constructed by averaging a signal with its shifted
version $\frac 12(s_n + s_{n-1})_{n \in \mathbb{Z}}$. Analogously, we
construct a low-pass filter by summing the trivial shift and all
shifts by $q$, $\tilde{H} = I + \sum_{q \in\vrt} T_q$, and normalize by
the largest eigenvalue: $H = 1/(|\lambda_{\max}|) \tilde{H}$.

In our example, the trivial filter $I$ can be written as
$I = -T_d + T_e + T_f$ and thus $H = T_a + T_b + T_c + 2 T_e + 2 T_f$ with coordinate vector
$\coord{h} = \tfrac{1}{6}(1,1,1,0,2,2)$. In matrix form it is
\begin{equation}
    \label{eq:SimpleLowPassFilter}
    H = \frac{1}{6}
	\left[\footnotesize\ra{1.0}
	\begin{array}{@{}rrrrrr@{}}
        6 & 0 & 0 & 0 & 0 & 0 \\
        0 & 6 & 0 & 0 & 0 & 0 \\
        0.9 & 0.6 & 3 & 0 & 0 & 0 \\
        1.4 & 1.4 & 0 & 4 & 0 & 0 \\
        1.6 & 1.3 & 1 & 1 & 2 & 0 \\
        0.7 & 1.1 & 0 & 1 & 0 & 2 
	\end{array}
	\right]. 
\end{equation}
The frequency response of this filter is $\fr{\coord{h}} = (1, 1, 1/2, 2/3, 1/3, 1/3)$ which shows that indeed higher frequencies (associated with later nodes in the DAG) are attenuated (the highest two by $1/3$), while, in this case, the lowest two are maintained.


\subsection{Relation to structural equation models}\label{linsemsec}

Structural equation models (SEMs), also called structural causal models (SCMs), are an important tool for analyzing causal data \cite{Peters:2017}, modeled as a DAG of causally dependent random variables that satisfy functional relationships. For the special class of linear SEMs (e.g.~\cite{Zheng:2018}, which uses them to learn DAGs from data), this relationship is linear. 

Assuming, as before, a weighted DAG $\dg = (\vrt,\edg, A)$ with $n$ nodes, a linear SEM  is defined as 
\begin{equation}\label{linsem}
X = AX + N,
\end{equation}
where $X = (X_1,\dots,X_n)^T$ is a random vector and $N = (N_1,\dots, N_n)^T$ a (usually i.i.d.) random noise vector, not necessarily Gaussian.  Using \eqref{eq:nicew} we can write \eqref{linsem} as
\begin{equation}\label{linsemsolved}
(\one_n - A)X = N\Leftrightarrow X = WN = \Fourier_\dg^{-1}N,
\end{equation}
if we choose the $(+,\cdot)$-transitive closure associated with the pollution model (but now without restriction on the weights) to obtain $W$ from $A$. In words, the noise chosen to sample the linear SEM is then, in our sense, the spectrum of the obtained signal.

Conversely, assume a weighted DAG $\dg = (\vrt,\edg, A)$ with $n$ nodes and an arbitrarily chosen weighted transitive closure $W$ (e.g., from Table~\ref{tab:ClosedSemirings}), which defines a notion of spectrum in the sense of this paper. Let $\dg'= (\vrt, \edg',A')$ be the DAG associated with $A'= \one_n - W^{-1}$. Then the linear SEM $X = A'X + N$ is such that the noise chosen to sample is the spectrum of the obtained signal.

The latter allows the interpretation of any weighted transitive closure in the context of linear SEMs.

\section{Discussion and Related Work}\label{sec:dis}

We discuss related work and some of the salient aspects of our causal SP framework for DAGs.

\mypar{Comparison to graph SP} Graph SP~\cite{Ortega.Frossard.Kovacevic.Moura.Vandergheynst:2018a} is concerned with signals indexed by the nodes of a graph and generalizes classical SP concepts by choosing adjacency matrix or Laplacian, or variants thereof as shift (or variation) operator~\cite{Sandryhaila:13,Shuman:13}, which is known to be the defining concept of any linear SP framework~\cite{Pueschel:08a}. For undirected graphs, the eigendecomposition of the shift exists and yields an orthogonal Fourier transform. Other SP concepts and techniques take meaningful forms~\cite{Ortega.Frossard.Kovacevic.Moura.Vandergheynst:2018a}. For directed graphs (digraphs) a proper generalization was still considered an open problem in~\cite[Sec.~III.A]{Ortega.Frossard.Kovacevic.Moura.Vandergheynst:2018a}, since an eigendecomposition does not exist in general and the more general Jordan normal form is not computable. DAGs constitute, in a sense, a worst case among digraphs since the adjacency shift has only one eigenvalue zero.

Several solutions have been proposed for digraphs. An overview including applications of digraph signal processing can be found in \cite{Marques.Segarra.Mateos:2020a}. One approach computes a Fourier basis that minimizes the sum of directed variations~\cite{Sardellitti.Barbarossa.DiLorenzo:2017a}, or evenly spreads them~\cite{Shafipour.Khodabakhsh.Mateos.Nikolova:2018a,Shafipour.Khodabakhsh.Mateos.Nikolova:2019a}. Others include changing the shift to the Hermitian Laplacian~\cite{Furutani.Shibahara.Akiyama.Hato.Aida:2019a}, using an approximation based on the Schur decomposition~\cite{Domingos.Moura:2020a}, or adding generalized boundary conditions, i.e., additional edges to the digraph~\cite{Seifert.Pueschel:2020a}. 

All these are fundamentally different from our approach which is applicable only to acyclic digraphs and based on a very different notion of shift and Fourier basis. The fundamental difference is best captured in the shift definition: graph shifts capture the neighbor structure, whereas our causal shifts capture the partial order 
structure provided by DAGs, directly manipulating causes, i.e., values of predecessors inserted to the DAG. As a result, all derived concepts differ substantially. In particular, in our framework DAGs first need to be transitively closed (and their are choices) and the exclusive dependency on predecessors motivated by causality makes shifts and Fourier transform triangular.

Our work can equivalently be interpreted as Fourier analysis for signals on weighted posets, i.e., with a weight assigned to each pair $(x,y)$ with $x < y$.

\mypar{Comparison to lattice SP} Our work substantially generalizes SP on meet/join lattices~\cite{Pueschel:2019a,Pueschel.Seifert.Wendler:2020a,Seifert.Wendler.Pueschel:2021a}, which, in turn, generalizes SP with set functions from \cite{Pueschel:2018a,Pueschel.Wendler:20}. These lattices are a special class of posets or DAGs, in which each two elements have a unique greatest lower bound, which yields the concise representation of the shift in \eqref{splatt}. This paper drops this condition, which makes it applicable to arbitrary posets and thus arbitrary DAGs. Further, and equally important, we allow for non-trivial weights and thus interpretations like distance or influence, which should considerably expand applicability.

\mypar{Causality and linear SEMs} Most closely related in causality research are linear SEMs as we formally explained in Section~\ref{linsemsec}. Typically, in SEMs, linear or not, the causes of a node are the parents because of the form in \eqref{linsem}. In this paper we use the term differently, namely for all predecessors of a node motivated by the form in \eqref{linsemsolved}.\footnote{Thus, in our follow-up work \cite{misiakos2023learning} we used the term root causes instead.}

Linear SEMs have been studied in \cite{loh2014equalerror,peters2014identifiability,honorio2017learning, aragam2015concave}. One question is identifiability of the data distribution, which depends on the distribution of $N$ in \eqref{linsem}. If $N$ is i.i.d. Gaussian, linear SEMs can express any $n$-variate distribution with a suitable DAG \cite{aragam2015concave}. In contrast, motivated by our proposed Fourier analysis, our assumption in the linear SEM experiment later in Section~\ref{synthlearn} is that $N$ is approximately sparse with random support.

An active research area is learning the DAG from data, with various approaches specifically targeting linear SEMs \cite{shimizu2006lingam,zheng2018notears,bello2022dagma,ng2020GOLEM}. In particular, \cite{zheng2018notears} captures acyclicity as a continuous constraint to obtain a solvable  optimization problem. Our work \cite{misiakos2023learning} builds on it but changes the data generation from \eqref{linsem} to assume sparsity in the Fourier domain. Thus we believe that our has the potential to bring new, SP-inspired method to the domain of causal data analytics. We also note that the interpretation of different transitive closures as linear SEMs (Section~\ref{linsemsec}) appears to be novel.


\mypar{Multiple shifts} Our framework is shift-invariant and based on multiple basic shifts instead of just one in graph SP. This is not uncommon: e.g., filters on images are composed from independent shifts in $x$ and $y$-direction, so also there the spectrum is partially ordered. Fundamentally, it just means that the filter space is a polynomial algebra in multiple variables~\cite{Pueschel:08a}.

\mypar{SP on non-Euclidean domains} Besides graph SP, other SP frameworks for non-Euclidean domains have been proposed. 

One line of work is topological SP~\cite{Barbarossa:20} based on the Hodge Laplacian, which considers signals defined on simplicial complexes, with values assigned to nodes, edges, or higher-order faces. The framework was generalized to cell complexes in \cite{Roddenberry.Schaub.Hajij:2021a,Sardellitti.Barbarossa.Testa:2021a}.

Hypergraphs generalize graphs by allowing edges with more than two nodes. A topological approach to hypergraph SP similar to above was proposed in~\cite{Barbarossa:16}, whereas~\cite{Zhang:20} uses the adjacency tensor and
tensor decomposition to define a notion of spectrum, sampling theory, and filters.

Quiver SP~\cite{Paradamayorga:20} considers directed multigraphs (i.e., multiple directed edges between the same nodes are possible), and develops an SP theory based on the rich representation theory of these structures.

Graphon SP is a continuous extension of SP on undirected graphs, thus enabling sampling among other things~\cite{Ruiz:21}.

SP on lattices and powersets \cite{Pueschel.Seifert.Wendler:2020a,Pueschel.Wendler:20} are direct predecessors of our work as already explained above and our first attempt to go to arbitrary unweighted DAGs was in \cite{Seifert.Wendler.Pueschel:2022a}.

Several of the above generalized SP frameworks, and the work in this
paper, build on the algebraic signal processing theory, which provides
the axioms, insights, and derivation guidelines for any linear SP
framework~\cite{Pueschel:08a,Pueschel:06c} and was used to consider
the first shifts beyond standard
translation/delay~\cite{Pueschel:08b,Pueschel:07,Sandryhaila:12}.

\section{Application: Fourier Sparsity}
\label{sec:ApplicationLinearSEM}%

Our work provides a complete set of basic SP concepts for data on weighted DAGs. Thus, in principle, any SP method that builds on Fourier analysis or filtering can be ported. In this paper we focus on the concept of sparsity in the Fourier domain, which, in the causal setting, has the appealing equivalent interpretation of signals with few causes.

As an example, consider again the river network with measured pollution data from Section~\ref{basic}. Fourier sparsity means that only few cities polluted in a data set, a reasonable assumption.

There are two problems that one can readily associate with Fourier sparsity:
\begin{itemize}
	\item Reconstructing a DAG signal from samples under the assumption of Fourier sparsity.
	\item Learning the DAG from DAG signals under the assumption of Fourier sparsity.
\end{itemize} 
We proposed a solution, called SparseRC, for the second problem in the follow-up work \cite{misiakos2023learning} for the special case of linear SEMs, i.e., the $(+,\cdot)$-transitive closure of the pollution model without weight restriction as explained in Section~\ref{linsemsec}.
The problem had not been considered before. We proved identifiability and showed that SparseRC  could successfully learn DAGs up to thousands of nodes under mild assumptions. Further, in the recent CausalBench challenge \cite{chevalley2022causalbench} on learning gene interactions from single cell data, SparseRC was among the three winning teams \cite{cbresults23}, showing that the assumption of few causes can be relevant in practice.

In this paper we focus on the first problem of reconstructing a signal from samples. First, as a proof of concept, we consider a synthetic problem on randomly generated graphs, again based on a linear SEM. Then, in the following Section~\ref{sec:ApplicationDynamicNetworks}, we consider a more realistic semi-synthetic experiment at a much larger scale modeling infection spreading on a dynamic network along time. In that experiment the assumption of Fourier-sparsity is intuitive but not explicitly built into the experiment.

\subsection{Learning DAG signals from samples}\label{synthlearn}

We generate signals on weighted, random DAGs using a corresponding linear SEM under the assumption of approximate Fourier-sparsity with unknown support w.r.t.~our associated Fourier basis. Then we reconstruct the signal from samples, using
a Lasso-method~\cite{tibshirani1996regression} (i.e., linear regression with a sparsity penalty) that is applicably with any basis.
We compare against prior graph Fourier bases obtained by dropping directions in the DAGs.
 
\mypar{Random graphs} We construct weighted DAGs $\dg = (\vrt,\edg,A)$ with $500$ nodes using the Erdős–Rényi model~\cite{erdos}, where each edge is created with probability $p =
0.05$ and given a random weight in $[-1,1]$. To obtain a DAG, we order the nodes randomly and only keep the edges $(y,x)$ with $x \geq y$. We
construct $100$ DAGs, each one thus with approximately $6000$ edges.

\mypar{Signal generation} We generate approximately Fourier-sparse signals on a random DAG using a linear SEM as described in~\cite{misiakos2023learning}. Namely, the inverse Fourier transform $\Fourier_\dg^{-1} = W$ is the $(+,\cdot)$-transitive closure (Section~\ref{linsemsec}) and the data is generated via
\begin{equation}
    \label{eq:LinearSEMSignalGeneration}
    X = \Fourier_\dg^{-1} (C + N_c) + N_x,
\end{equation}
where $C$ is the sparse spectrum, i.e., the relevant few causes, $N_c$ is spectral noise, and $N_x$ models the noise in the
measurement of $X$. Here the $C + N_c$ term corresponds to the $N$ term
in~\eqref{linsemsolved}. Both $N_c$ and $N_x$ are assumed to be of negligible magnitude compared to $C$. Specifically, we choose $C$ to be sparse with only $10\%$ nonzero values at random locations and in the range $[1,10]$, and $N_c$ and $N_x$ are i.i.d.~zero-mean Gaussian noise with a standard deviation of $0.1$.

In the river network example, $C$ would be the inserted pollution by a city, $N_c$ negligible random pollution inserted by all cities, and $N_x$ the noise in the pollution measurement.

\mypar{Reconstruction method} We reconstruct the generated signal $\coord{s}$ from samples $s_{x_i}$ at nodes $x_1,\dots,x_k$ by fitting a Lasso model. That is, we solve  a linear regression with an $L_1$-sparsity penalty:
\begin{equation}
    \label{eq:LassoModel}
    \min_{\ft{\coord{r}} \in \mathbb{R}^{|V|}}
    \sum_{i=1}^k (s_{x_i} - \sum_{y \in\vrt} \ft{r}_y f^y_{x_i} ) ^2 + \lambda \|  \ft{\coord{r}} \|_1. 
\end{equation}
Here $\coord{f}^y$ is the $y$-th Fourier basis vector \eqref{eq:FourierBasis}, the linear regression approximates the samples $s_{x_i}$
as a signal on the DAG, while the $L_1$-penalty  promotes sparsity in the Fourier spectrum. With the minimizing $\ft{\coord{r}}$ found the obtained reconstructed signal is then
$$
\coord{r}= \Fourier_\dg^{-1}\ft{\coord{r}}.
$$

\mypar{Baselines} As baselines we consider other Fourier bases in \eqref{eq:LassoModel}. Namely, the graph Fourier bases associated with adjacency and Laplacian matrices, for both the DAG and its transitive closure obtained by dropping the directions in the DAG, i.e., the eigenbases of $A+A^T$ and $\cl{A} + \cl{A}^T = W+W^T-2I$.

Further, we also consider our binary DAG Fourier basis obtained by setting all nonzero weights in $A$ to one and computing the Boolean transitive closure.

\begin{figure}
	\centering
	\includegraphics[width=\linewidth]{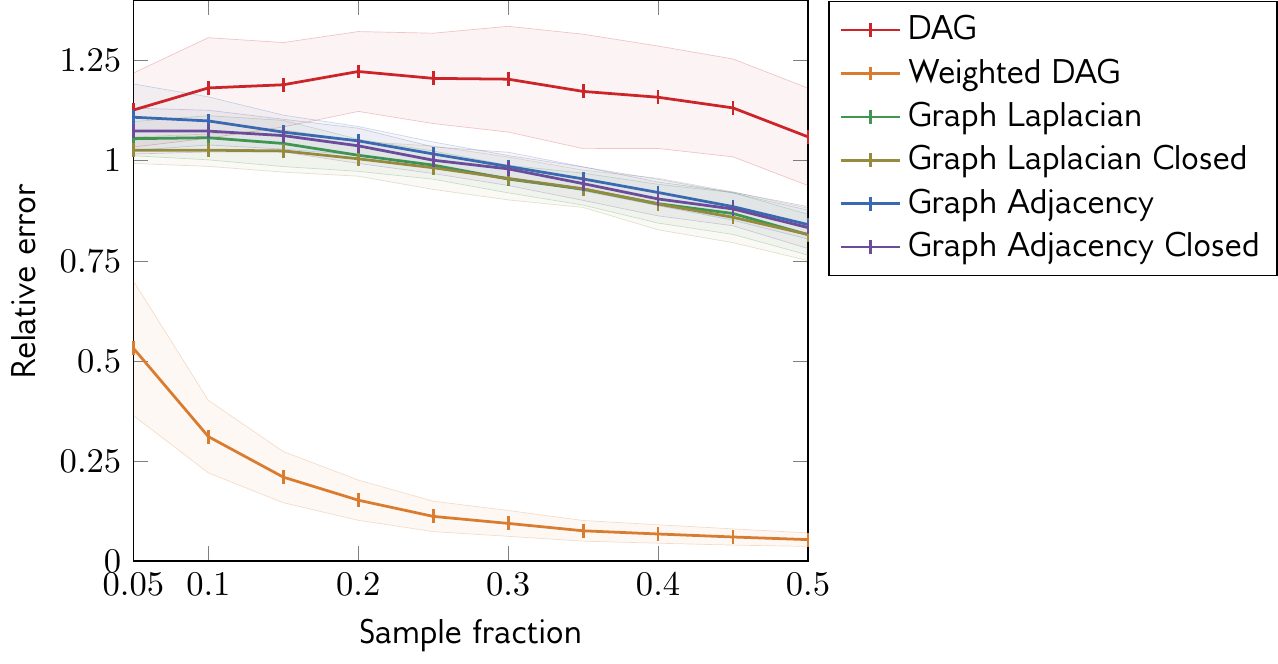} 
	\caption{Relative error of reconstructed signal.}
	\label{fig:LinearSEMSignalRelativeError}
\end{figure}

\mypar{Results} Fig.~\ref{fig:LinearSEMSignalRelativeError} shows the results: the reconstruction error
$\norm{\coord{r} - \coord{s}} / \norm{\coord{s}}$ as a function of the fraction of the signal sampled. The shaded areas show the $95\%$ confidence intervals over 100 repetitions.

As expected, given enough samples, the signal can be well reconstructed with the Fourier basis used in its construction. The other bases fail since the signal is not approximately sparse in their Fourier domain. This also applies to our unweighted DAG basis. In other words, the weights and chosen transitive closure matter.

\section{Application Example: Dynamic Networks}
\label{sec:ApplicationDynamicNetworks}%

We present a more realistic example of learning a DAG signal from samples, again under the assumption of Fourier sparsity, i.e., few causes, but this time this sparsity is not present by construction. As signal domain we consider a class of DAGs that is obtained from graphs that change dynamically with time. 

Examples of real-world (undirected) graphs  include proximity of persons (used, e.g., for contact-tracing of infectious people during a pandemic), peer-to-peer networks between vehicles in traffic, or transactions between traders in a market. However, these graphs are often non-static, e.g., the edges change with time. The work in~\cite{Kim.Anderson:2012a} shows how to encode such a dynamic network as a DAG by assigning the graphs to discrete time steps and connecting subsequent graphs. 

We consider such DAGs as one possible application domain of our work and present in this section a prototypical, semi-synthetic example: we use real contact-tracing data to simulate the spread of an infection among $n$ individuals along time. Then we try to reconstruct, or learn, this infection signal from samples under the assumption of sparsity in the Fourier domain. We presented a restricted, simplified version of this experiment using unweighted DAGs in~\cite{Seifert.Wendler.Pueschel:2022a}.

\subsection{Infection Spreading on Dynamic DAGs}
\label{subsec:DynamicNetworkDAGs}%

We explain the construction of DAGs from dynamically changing graphs and the model we use to generate infection signals from contact tracing data.

\mypar{Dynamic networks as DAGs} We consider a dynamic network as a collection of (undirected) graphs $\graph_t = (\vrt,\edg_t)$ where the set of edges $\edg_t$ changes with time
$t \in {\cal T} = \{t_1,\dots, t_m\}$. It can be modeled as a DAG
$\dg = (\vrt', \edg')$ using the idea from~\cite{Kim.Anderson:2012a}. Namely, 
we make a copy of the node set for each time point, i.e., the new node set is
$\vrt'= \{ (v,t) \; | \; v \in\vrt, t \in {\cal T} \cup \{t_{m+1}\}\}$, where $t_{m+1}$ is an added, last time point.

\begin{figure*}
	\centering
	\hfill
	\subcaptionbox{A dynamic network $(\vrt,\edg_t)$, where the edges change with
		time $t=t_1,t_2,t_3$. \label{subfix:DynamicNetwork}}[0.225\linewidth] { 
		\includegraphics[width=\linewidth]{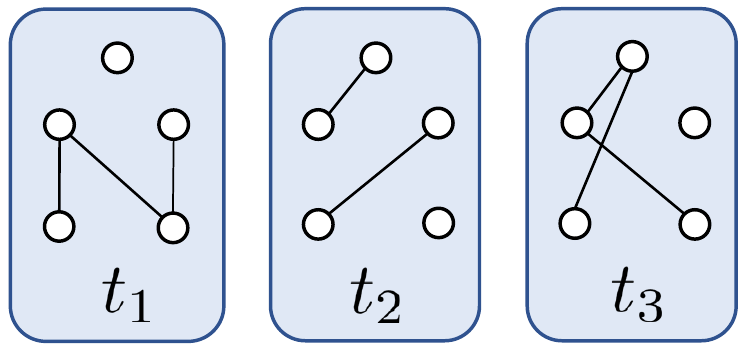}
	}
	\hfill
	\subcaptionbox{Copied graphs with new directed edges. The
		edges $(u,t) \to (u,t+1)$ are not yet included. \label{subfix:DynamicNetworkDAG}}[0.3\linewidth] {
		\includegraphics[width=\linewidth]{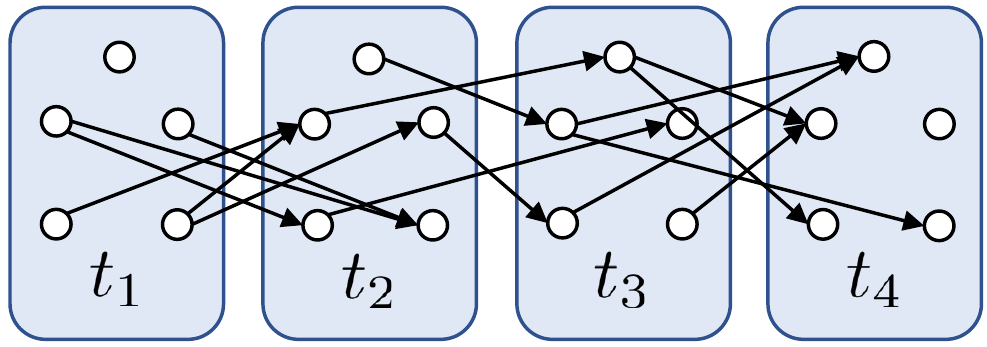} }
	\hfill
	\subcaptionbox{The final DAG $\dg = (\vrt',\edg')$; the nodes at each time
		step are drawn vertically
		aligned. \label{subfix:DynamicNetworkDAGFinal}}[0.3\linewidth]
	{
		\hspace*{\fill}%
		\includegraphics[width=0.5\linewidth]{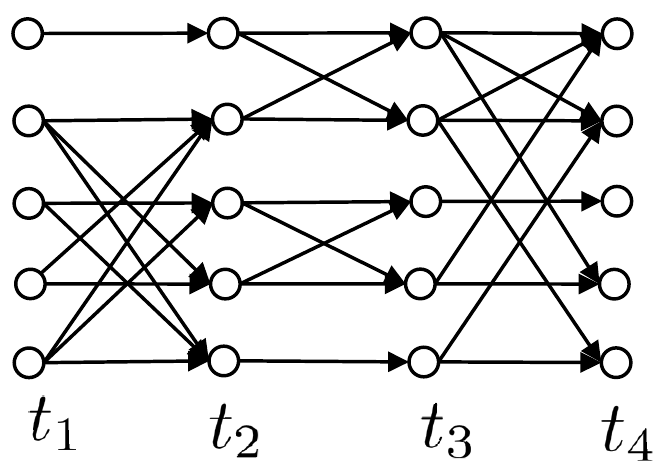}
		\hspace*{\fill}%
	}
	\hfill    
	\caption{Constructing a DAG from a dynamic network~\cite{Kim.Anderson:2012a}.}
	\label{fig:DynamicNetworkToDAG}
\end{figure*}

Further, we connect nodes $(u,t)$ with $(v,t+1)$ if $(u,v)\in \edg_t$ and always
$(u,t)$ with $(u,t+1)$ to form $\edg'$. This construction is illustrated
on a small example in Fig.~\ref{fig:DynamicNetworkToDAG}.

\mypar{The Haslemere data set} We consider the data set from
\cite{Kissler.Klepac.Tang.Conlan.Gog:2018a}, which uses real smartphone proximity data
to obtain a dynamic network on which the spread of a disease is then modeled and analyzed. Here we aim to learn the associated signal from samples. Concretely, the proximity of $|\vrt| = 469$ participants was measured for three days every 5 minutes between 7am and 11pm using a smartphone app, resulting in $576$ time points. Due to our infection model below we remove all
edges with distance $> 20$ meters.

\mypar{Haslemere DAG} With the above construction we turn the Haslemere dynamic network into a DAG $\dg = (\vrt',\edg')$. Since later we want to use standard graph SP as benchmark (with directions dropped), which requires an eigendecomposition of the Laplacian or adjacency matrix, we have to
restrain the size of $\dg$. Thus, we sample the contact data only
every hour, resulting in $|{\cal T}| = 37$ time points leading to a DAG with
$|\vrt'| = 17612$ nodes and $|\edg'| = 24596$ edges.

\mypar{Weights} At each time point the edges of the graphs $G_t$ are weighted by the
distance of the participants. We convert the distances into influences as explained in Section~\ref{sec:wtc} to ensure fading with large distances in space and, through the transitive closure, in time. Concretely, DAG edges of the form $((u,t), (u,t+1))$ obtain weight 1, and edges of the form $((u,t), (v,t+1))$ obtain weight $e^{- d_{u,v}} \in [0,1]$, where $d(u,v)$ is the distance at time $t$. The transitive closure is then computed using Algorithm~\ref{algo:ModifiedFloydWarshall}.

\mypar{Infection signals} \cite{Kissler.Klepac.Tang.Conlan.Gog:2018a} uses the susceptible-exposed-infectious (SEI) model to simulate the spread of a disease from a number of initially infected individuals. In this model, a healthy individual is infected with a certain probability when exposed to an infected individual. Here, to make it more realistic, we include recovery and slightly extend it to a susceptible-infected-recovered (SIR) model.

Formally, from~\cite{Kissler.Klepac.Tang.Conlan.Gog:2018a}, the infection force $\lambda_{u,v}(t)$ from an infected individual (node) $u$ to a non-infected individual (node) $v$ at each time point is modeled using a cutoff exponential
\begin{equation}\label{eq:HaslemereInfectionForce}
    \lambda_{u,v}(t) =
    \begin{cases}
        \E^{-d_{u,v}(t)/\rho} & \text{if } d_{u,v}(t) \leq \epsilon, \\
        0 & \text{if } d_{u,v}(t) > \epsilon,
    \end{cases}
\end{equation}
where $d_{u,v}(t)$ is the distance between $u$ and $v$ at time $t$,
$\rho$ the characteristic distance set to $\rho = 10$ meters, and
$\epsilon$ the cutoff distance set to $\epsilon = 20$ meters. The
overall infection force to a node $v$ is then
\begin{equation}
    \label{eq:FullInfectionForce}
    \lambda_v(t) = \sum_{u \text{ infected}} \lambda_{u,v}(t),
\end{equation}
and the probability that the individual $v$ gets infected at time $t$ is
\begin{equation}
    \label{eq:InfectionProbability}
    \prob_v(t) = 1 - \E^{-\lambda_v(t)}.
\end{equation}
In our extension, a person which is infected at time $t$ recovers at time $t+5$ and is
afterwards immune. The time of 5 hours is of course unrealistically short, but this is necessary due to the small number of considered time points. The exact choice is also irrelevant for our prototypical experiment.

\subsection{Learning Fourier-Sparse Causal Signals}
\label{subsec:LearningFourierSparse}%

\begin{figure}
    \centering
    \includegraphics[width=0.75\linewidth]{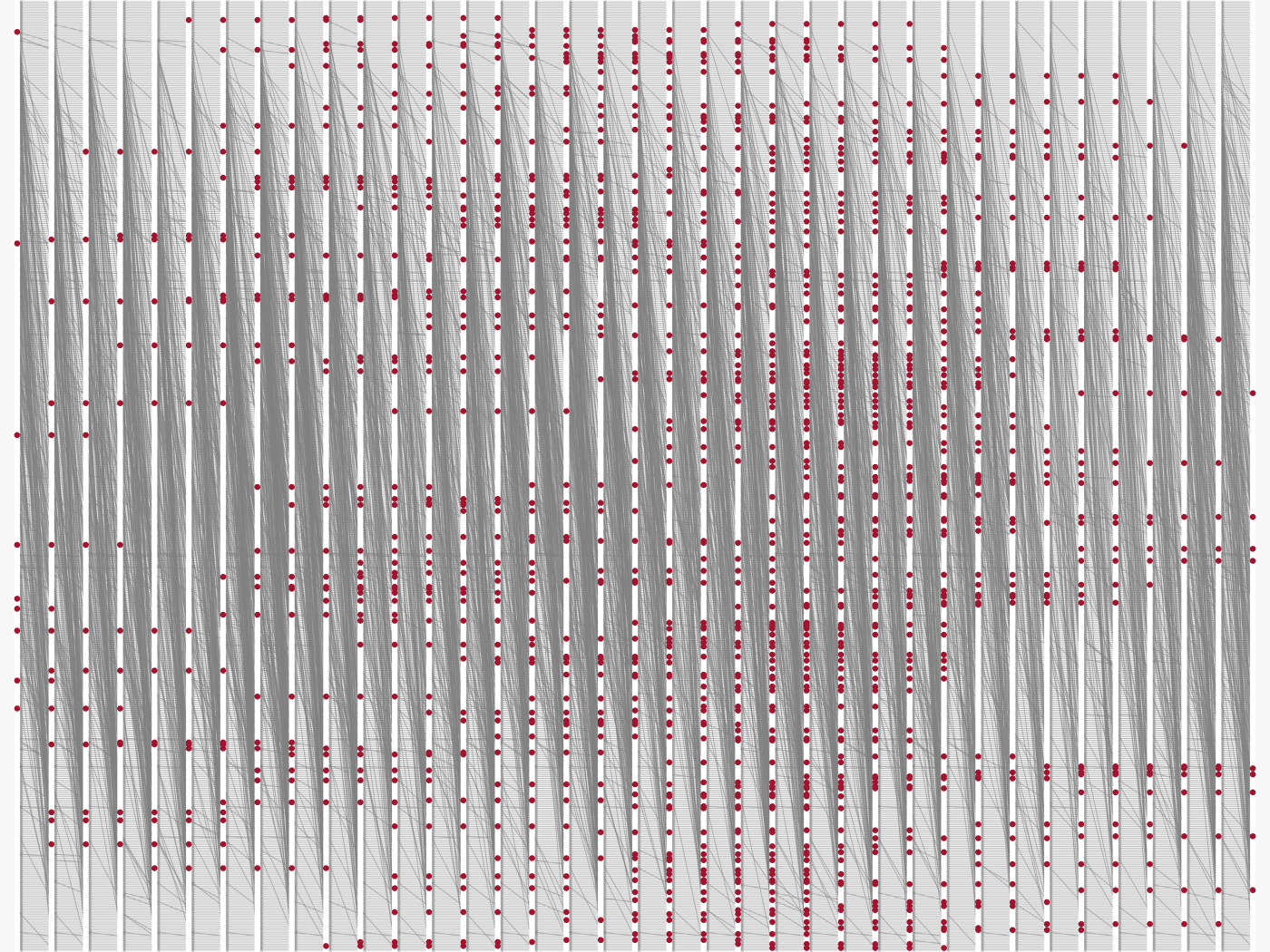} 
    \caption{One example of a generated binary infection signal with nine initially infected persons. About 9\% of the values are $=1$ (infected), the others $=0$ (not infected).}
    \label{fig:Signal9InfectedPersons}
\end{figure}

Using the model from Section~\ref{subsec:DynamicNetworkDAGs} we can
generate binary (values are 0 or 1) infection signals on the DAG
$(\vrt', \edg')$ by starting with a small number of infected individuals at
time point one and infecting individuals in subsequent time steps with
the probabilities \eqref{eq:InfectionProbability}. Fig.~\ref{fig:Signal9InfectedPersons} shows one such signal with nine initially infected individuals. Since the
connectivity information (i.e., the edges of the DAG) is distracting
from the signal (the infected individuals, i.e., the red dots) we will omit the edges in the following plots. Fig.~\ref{fig:Signal9InfectedPersons} without edges is shown later in Fig.~\ref{ex:ExampleSignal} where it is compared to its reconstructions.

\begin{figure}
    \centering
    \includegraphics[width=0.75\linewidth]{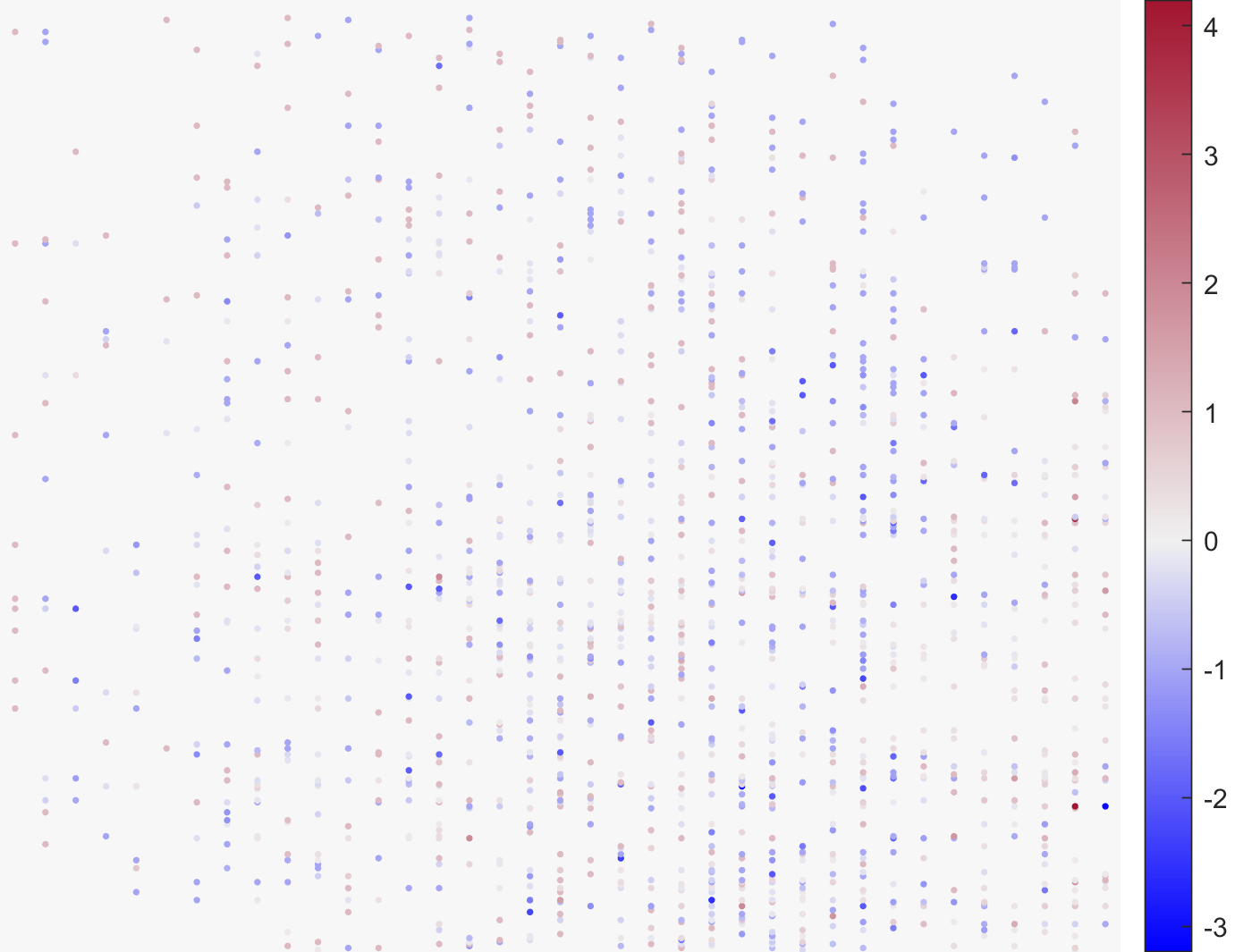} 
    \caption{Spectrum of the signal in Fig.~\ref{fig:Signal9InfectedPersons}. About 11\% of the values are nonzero, several of which are very small.}
    \label{fig:SignalInfected9PersonsSpectrum}
\end{figure}

Fig.~\ref{fig:SignalInfected9PersonsSpectrum} shows the spectrum of the signal (indexed by the same DAG because of Theorem~\ref{thm:PropsTotalVariation}, but now with edges omitted) in Fig.~\ref{fig:Signal9InfectedPersons}.

Now, we the goal is again to learn such a  generated infection signal from
a number of samples assuming sparsity in the Fourier domain. The approach is in concept similar to the one in Section~\ref{synthlearn}, but the details are different. We explain it next and then show results, again comparing to standard graph SP Fourier bases obtained by dropping the direction of the DAG edges so they are well-defined.

Learning such signals from samples is hard. First, the signals are very sparse; thus a certain number of samples is needed to learn something about the signal at all. Second, the DAG model does not know the data generation process. In particular, the fixed recovery time is not known or used. Finally, the data generation process is stochastic in nature and hence no model can be absolute certain about the infection status after exposure and along time.

\mypar{Fourier-sparse learning} The basic idea is to approximate a
binary infection signal $\coord{s}$ with $\tau(\sigma(\coord{r}))$, where $\coord{r}$ is Fourier-sparse (i.e, $\ft{\coord{r}}$ has few nonzero values), $\sigma = 1/(1+\E^{-x})$ is the
logistic sigmoid function that converts elementwise real values to probabilities, and $\tau$ is a
threshold function. In our case, the default $\tau$ simply rounds elementwise to 0 or 1.

Formally, with this approximation, the probability that a node $x \in\vrt'$ is of class $1$ (infected) is then
\begin{equation}\label{eq:InfectionProbability1}
    p(x) 
    = \prob(x \text{ of class } 1)
    = \sigma(r_x)
    = \sigma\Big( \sum_{y \in\vrt} \ft{r}_y f^y_x \Big),
\end{equation}
with $f^y_x$ from \eqref{eq:FourierBasis} and $\ft{r}_y = 0$ for most $y \in\vrt$.

We assume we observe $k$ signal values $s_1,\dots,s_k$ at random nodes
$x_1,\dots,x_k$, respectively. We estimate the nonzero Fourier coefficients
$\ft{r}_y$ in \eqref{eq:InfectionProbability1} by solving a logistic regression problem, regularized by an $L^1$-loss term to promote sparsity of $\ft{\coord{r}}$~\cite{Ng:2004a}. The resulting optimization problem is given as
\begin{multline}
    \label{eq:LogisticRegressionProblem}
    \min_{\ft{\coord{r}} \in \mathbb{R}^{|V|}}
    - \sum_{i = 1}^k s_i \log p(x_i) + (1 - s_i) \log (1 - p(x_i))\\
    + \lambda \| \ft{\coord{r}} \|_1,
\end{multline}
where $\lambda \ll 1$ is a hyperparameter. We found that
$\lambda = 0.1$ worked well for all bases.

\mypar{Experiment} We generate a set of signals as follows. We start
the SIR model with $i = 5,9,11$ participants infected at random at
time $t_1$ and propagate the infections as described in Section~\ref{subsec:DynamicNetworkDAGs}. For each $i$ we repeated the simulation ten times, resulting in $30$ DAG signals overall. The
signals have value $1$ at node $x = (u,t)$ if the individual $u$ is
infected at time $t$ and value $0$ otherwise.

We observe a fraction $k/|\vrt'|$ of the signal values and then
use~\eqref{eq:LogisticRegressionProblem} with three different notions of Fourier
basis to obtain a sparse $\ft{\coord{r}}$, which in turn determines
$\coord{r}$ and thus the prediction $\tau(\sigma(\coord{r}))$ of
$\coord{s}$. For each of the three Fourier bases we consider two variants for a total of six experiments.

The three bases are our proposed basis and Laplacian/adjacency matrix GSP bases, as in Section~\ref{synthlearn}, obtained by dropping direction in the DAG $(\vrt',\edg')$ and computing eigenbases. For our basis we consider the weighted version, obtained by the transitive closure in the influence model as explained above, but also an unweighted version, obtained by a standard transitive closure (first row in Table~\ref{tab:ClosedSemirings}). For the Laplacian/adjacency matrix GSP bases we consider both the graph as is (but undirected) and its transitive closure.

All results shown are for the 30 considered signals: the respective solid lines show the mean and the shaded areas the $95\%$ confidence interval.

\begin{figure}
	\centering
	\includegraphics[width=\linewidth]{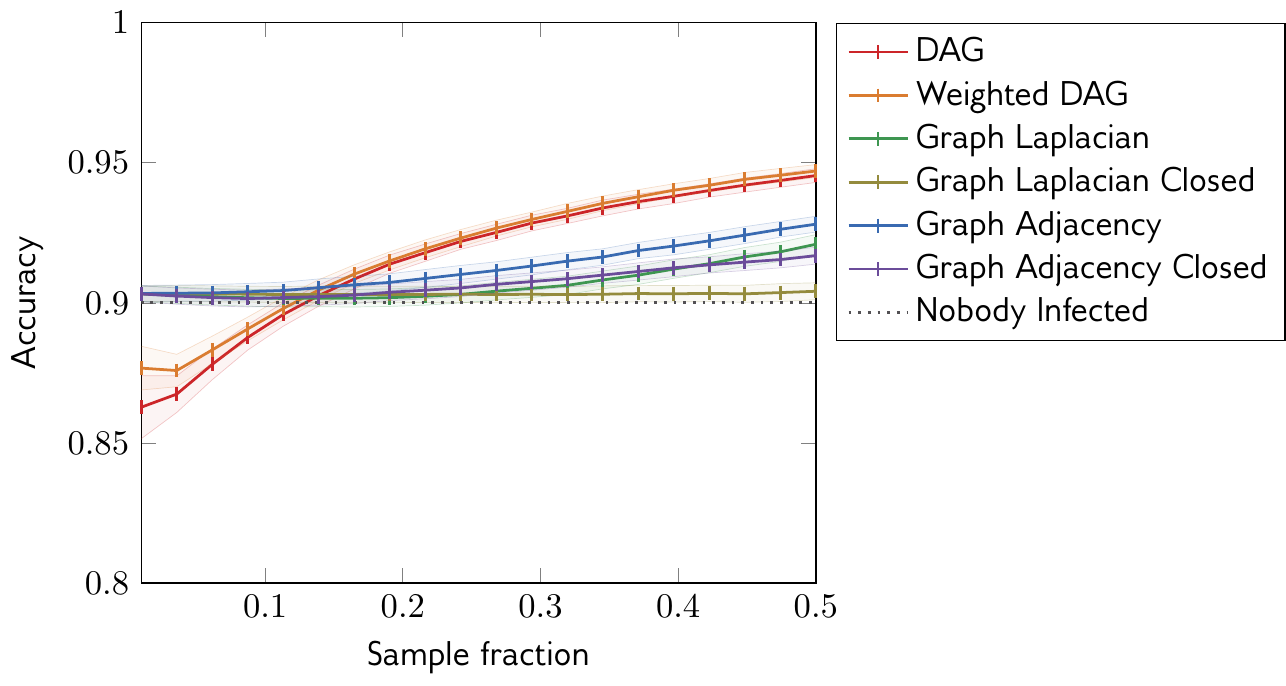} 
	\caption{Standard accuracy in the Euclidean norm is not a good measure for the quality of binary classifiers with imbalanced classes.}
	\label{fig:AccuracyPlots}
\end{figure}

\mypar{Evaluation} The first idea is to compute reconstruction accuracy, computed as $1 - ||(\coord{s} - \tau(\sigma(\coord{r}))||_2/||\coord{s}||_2$, shown in Fig.~\ref{fig:AccuracyPlots} (note that the $y$-axis starts at 0.8, which emphasizes differences). Since the signals are binary and highly imbalanced (way more person-time combinations are non-infected than infected), this metric is not suitable: a trivial estimator setting every value to ``nobody infected'' reaches about 0.9, shown as dotted line, but cannot detect any infected node and is thus useless. More generally, binary classifiers of similar such accuracy can have vastly different quality due to differences in the number of false positives.

Thus, instead, the quality of binary classifiers with imbalanced data is measured in machine learning with the \emph{receiver operator characteristic area under curve} (ROC-AUC)~\cite{Bradley:1997}. The key underlying concept is the ROC that does a cost/benefit analysis \cite{Fawcett:2006}. Namely, the so-called ROC curve measures how the true positive rate or TPR $y$ (the probability of detecting an event, i.e., the benefit) changes with respect to the false positive rate or FPR $x$ (the probability of a false alarm, i.e., the cost) by varying the classification threshold $\tau$. A random classifier
that chooses detection with probability $p$ and not detected with probability $1-p$ yields the ROC curve $y = x$ if $p$ is varied in $[0,1]$. A perfect classifier would approach the curve $y = 1$. So higher is better and the area under the ROC curve (ROC-AUC) is used as metric. The perfect classifier has AUC 1 and the trivial one (``nobody infected'') AUC 0.5. 

\begin{figure}
	\centering
	\includegraphics[width=\linewidth]{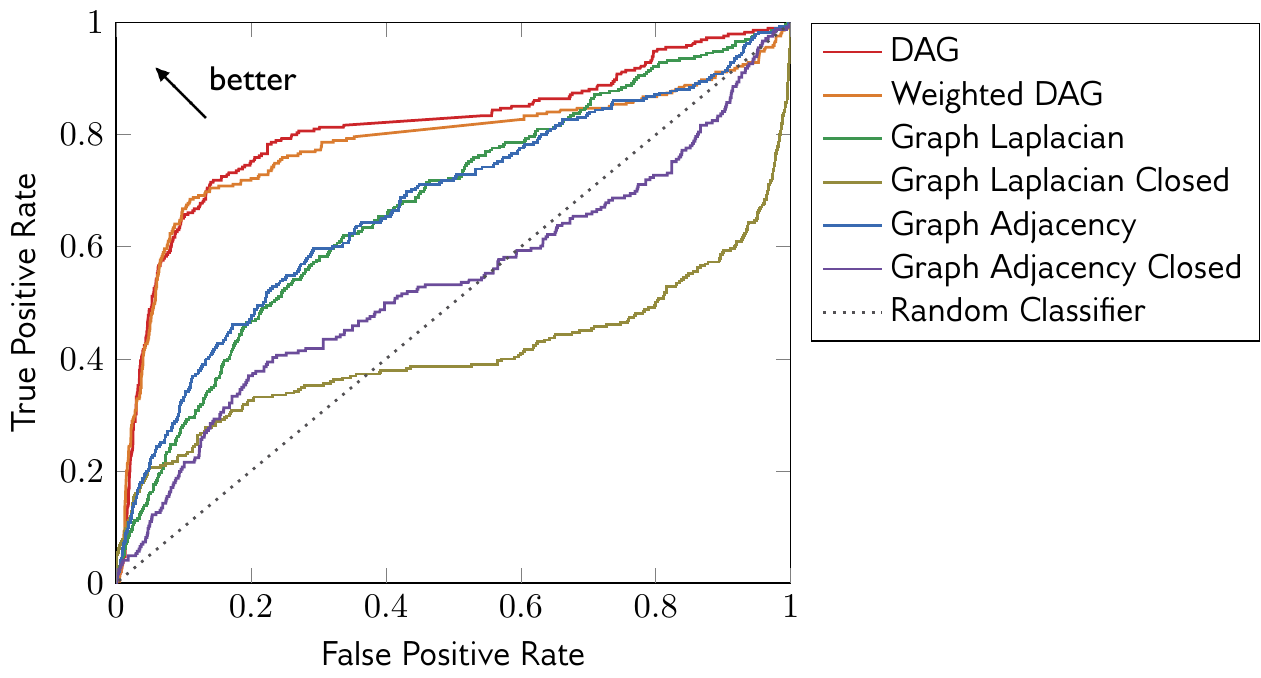} 
	\caption{Receiver operation characteristics (ROC) curves for the above
		sample data and the classifiers with a sample of $20\%$ node
		data.}
	\label{fig:ROCCurves9Persons}
\end{figure}

Fig.~\ref{fig:ROCCurves9Persons} shows the ROC curves of the considered
classifiers obtained with $20\%$ sampled data. The estimation based on our proposed causal Fourier basis performs best by a large margin, compared to both GSP Fourier bases associated with the Laplacian and adjacency matrix. Transitively closing the graphs makes it worse for them. The likely reason is that the for the GSP models the assumption of sparsity in the Fourier domain does not hold, and thus the signal could not be learned from samples using \eqref{eq:LogisticRegressionProblem}.

In contrast, in our Fourier domain sparsity indeed holds as already shown in Fig.~\ref{fig:SignalInfected9PersonsSpectrum}. In our terminology this means relatively few causes are responsible for the signal, which the Fourier-sparse reconstruction can leverage. It shows that for the considered signals our combinatorial causal Fourier basis yields a better representation than the more geometric GSP bases.

It is interesting that both considered DAG Fourier bases perform well in this experiment. There are two possible explanations. First, both provide a model for the data in which approximate Fourier sparsity holds. Second, the weighted model may be conceptually the better fit, but the binary nature of the data gives an advantage to the unweighted model since the associated Fourier basis consists of binary vectors as well.

\begin{figure}
	\centering
	\includegraphics[width=0.85\linewidth]{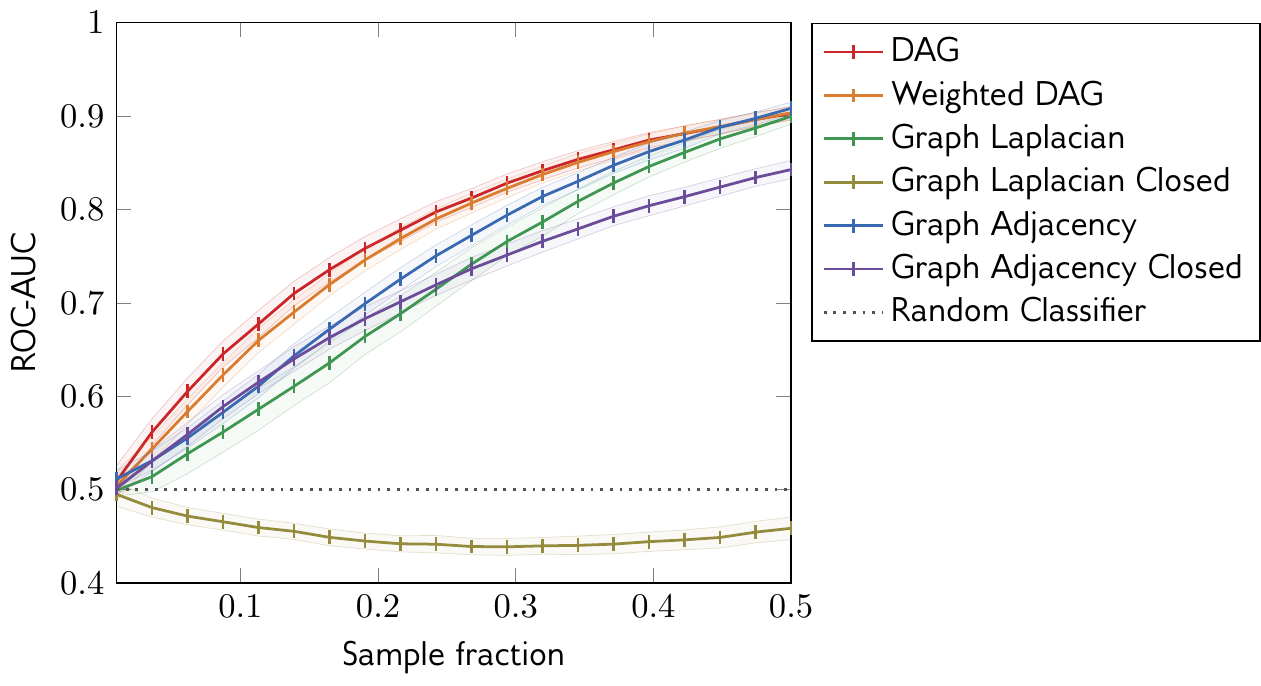}
	\caption{Results of the proposed sparse learning approach on the
		Haslemere signals in the ROC-AUC metric as a function of the sample fraction.}
	\label{fig:HaslemereResults}
\end{figure}

The corresponding ROC-AUC curve used to measure the benefit in ROC curves is shown in Fig.~\ref{fig:HaslemereResults}. It plots the AUC of the ROC lines as function of the fraction of data points sampled (higher is better). For the sample fraction of $20\%$ the values correspond to Fig.~\ref{fig:ROCCurves9Persons}.

\begin{figure*}
	\centering
	\subcaptionbox{Adjacency \label{ex:ExampleSignal}}[0.35\linewidth]
	{\includegraphics[width=1\linewidth]{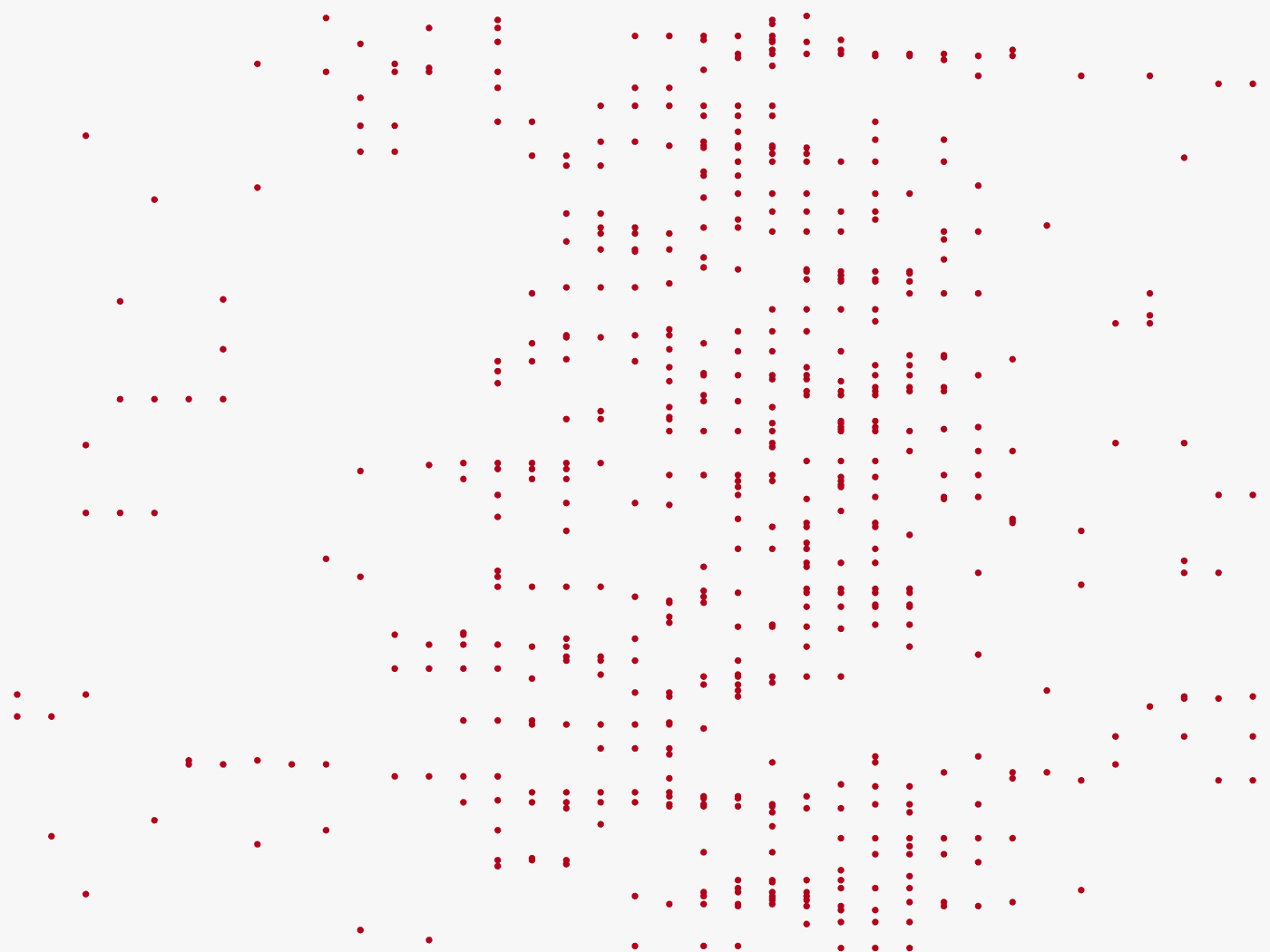}}
	\hspace{0.1\linewidth}
	\subcaptionbox{Adjacency closed \label{ex:ExampleSignal}}[0.35\linewidth]
	{\includegraphics[width=1\linewidth]{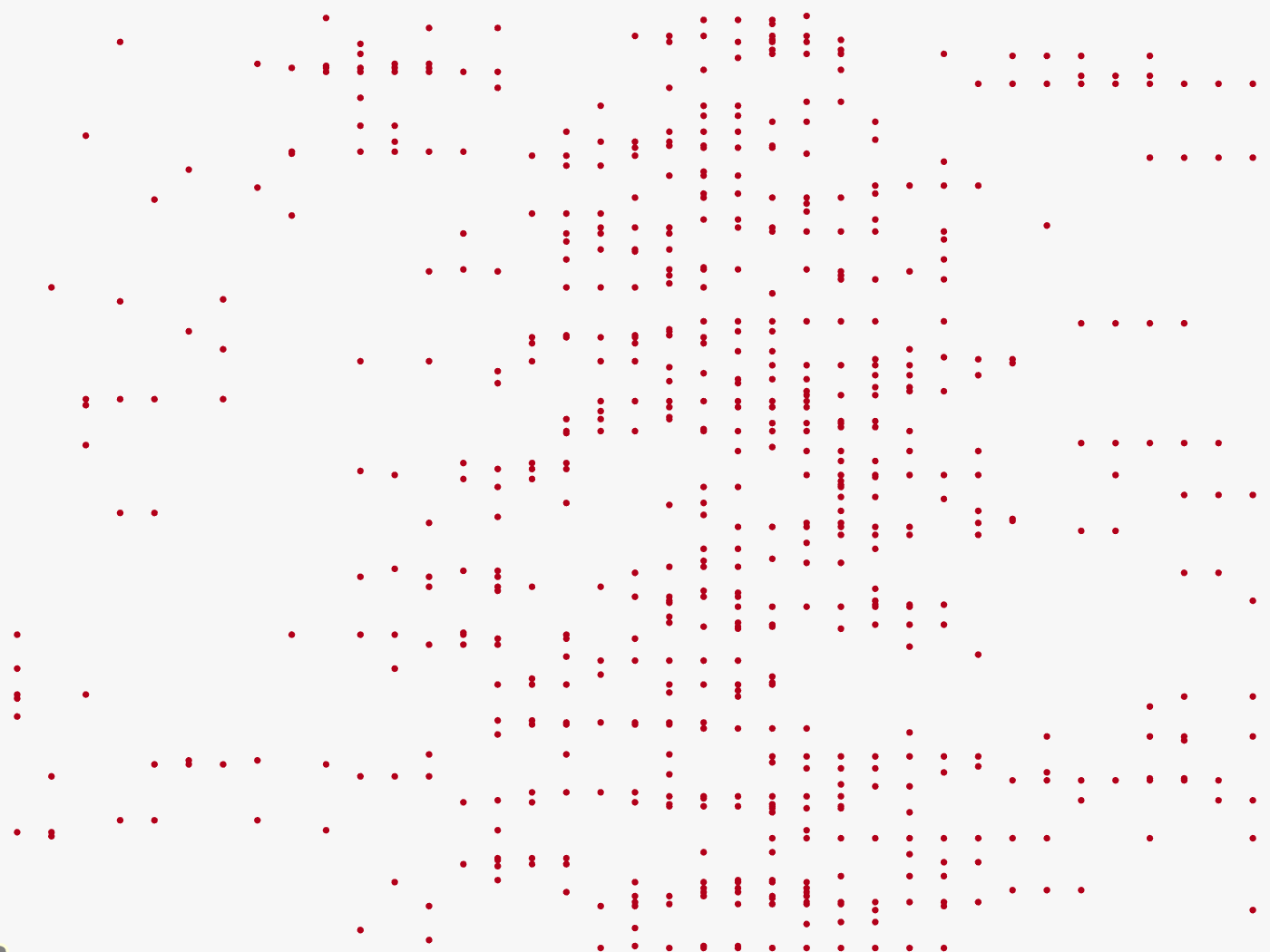}}
	\hspace{0.1\linewidth}
	\subcaptionbox{Laplacian \label{ex:ExampleSignal}}[0.35\linewidth]
	{\includegraphics[width=1\linewidth]{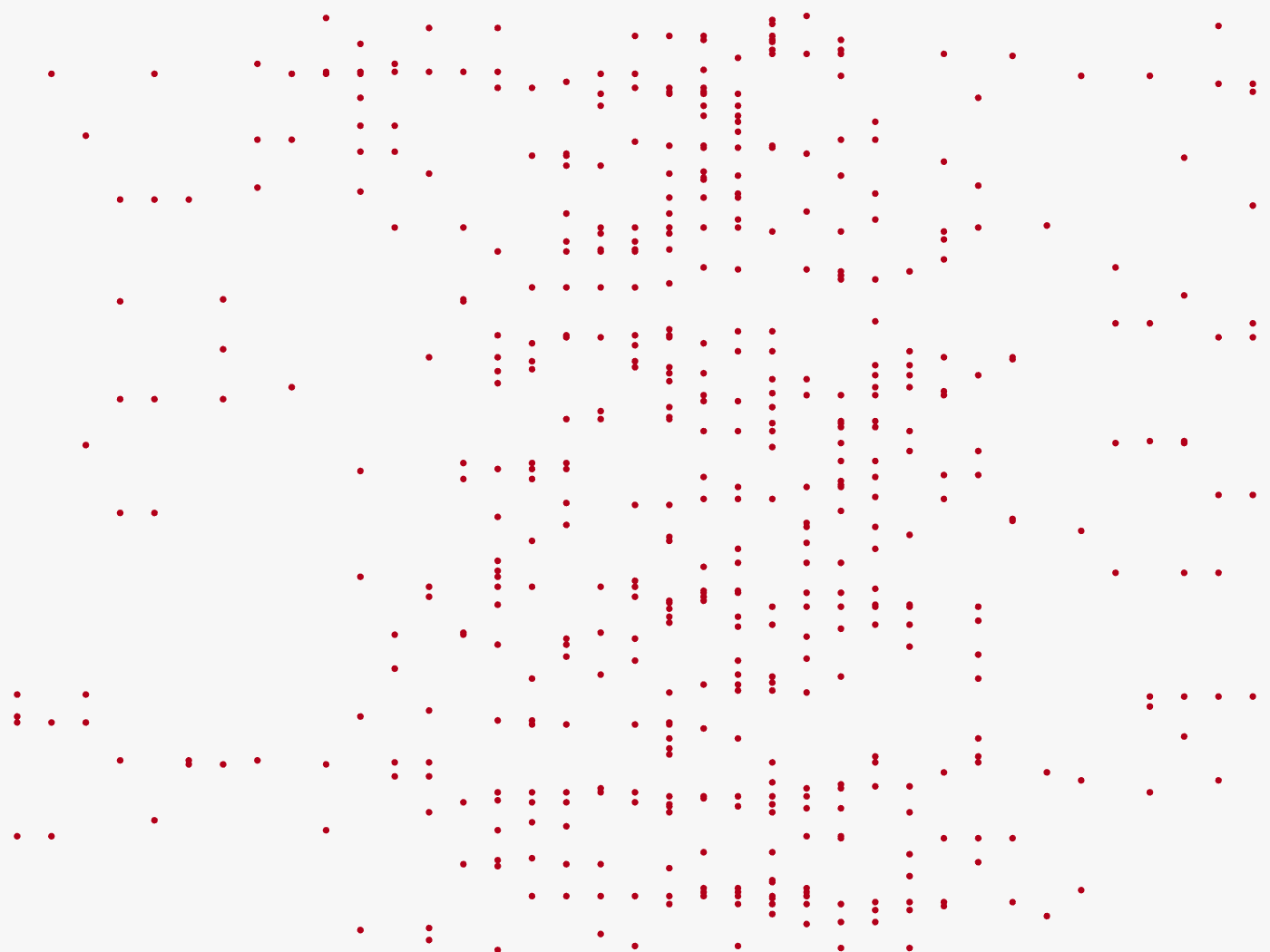}}
	\hspace{0.1\linewidth}
	\subcaptionbox{Laplacian closed \label{ex:ExampleSignal}}[0.35\linewidth]
	{\includegraphics[width=1\linewidth]{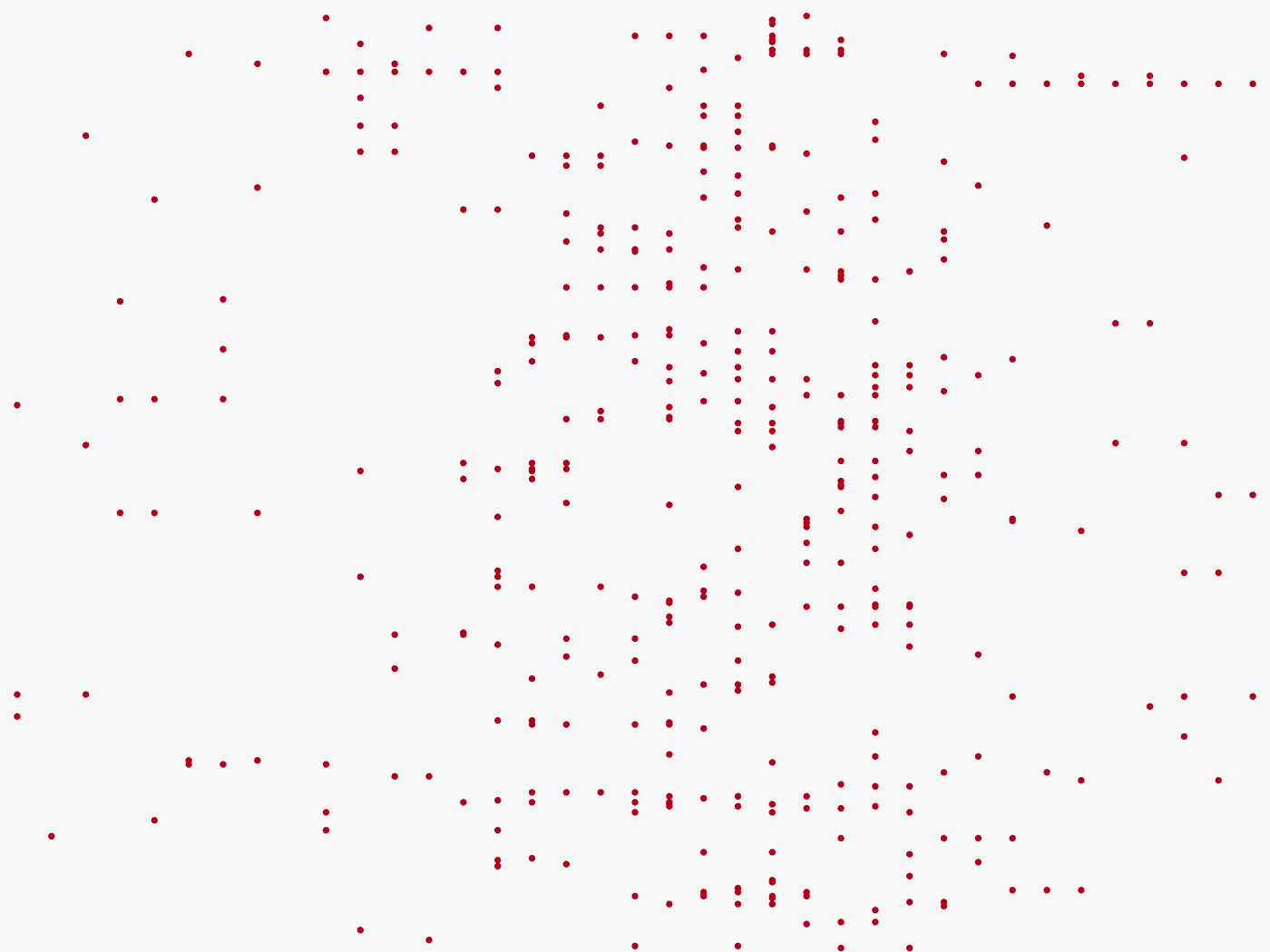}}
	\hspace{0.1\linewidth}
	\subcaptionbox{DAG \label{ex:ExampleSignal}}[0.35\linewidth]
	{\includegraphics[width=1\linewidth]{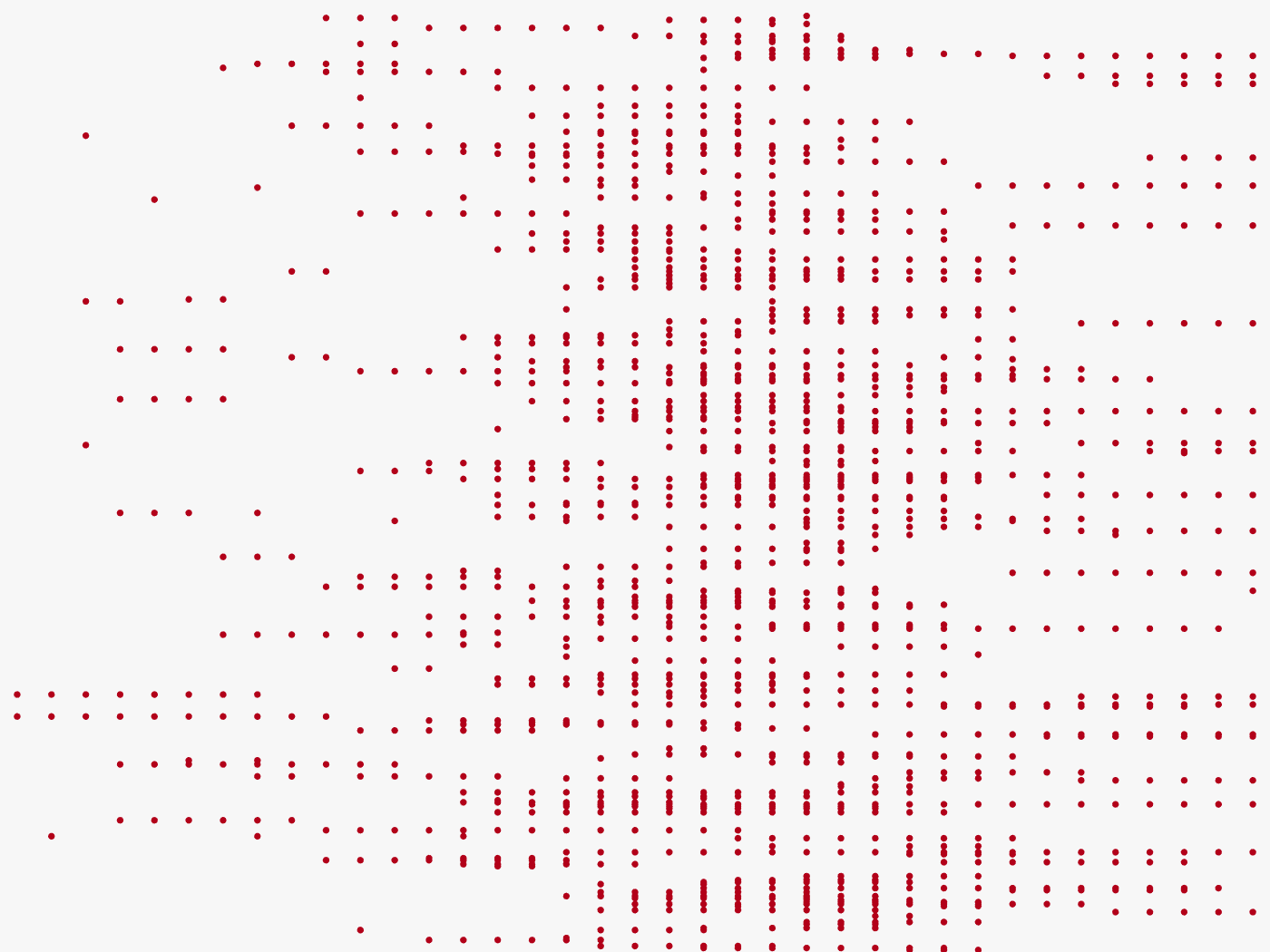}}
	\hspace{0.1\linewidth}
	\subcaptionbox{Weighted DAG \label{ex:ExampleSignal}}[0.35\linewidth]
	{\includegraphics[width=1\linewidth]{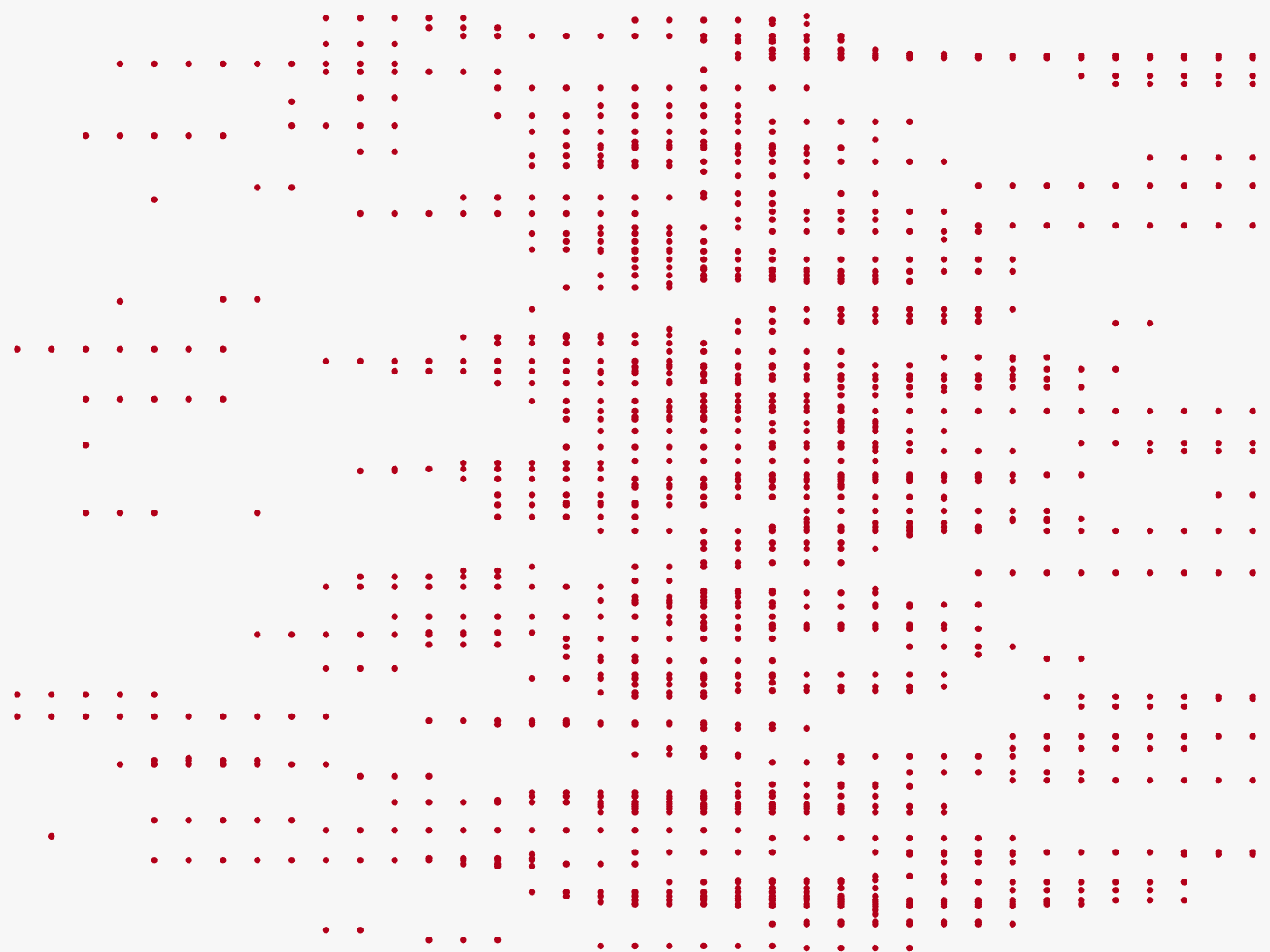}}
	\subcaptionbox{Original signal \label{ex:ExampleSignal}}[0.35\linewidth]
	{\includegraphics[width=1\linewidth]{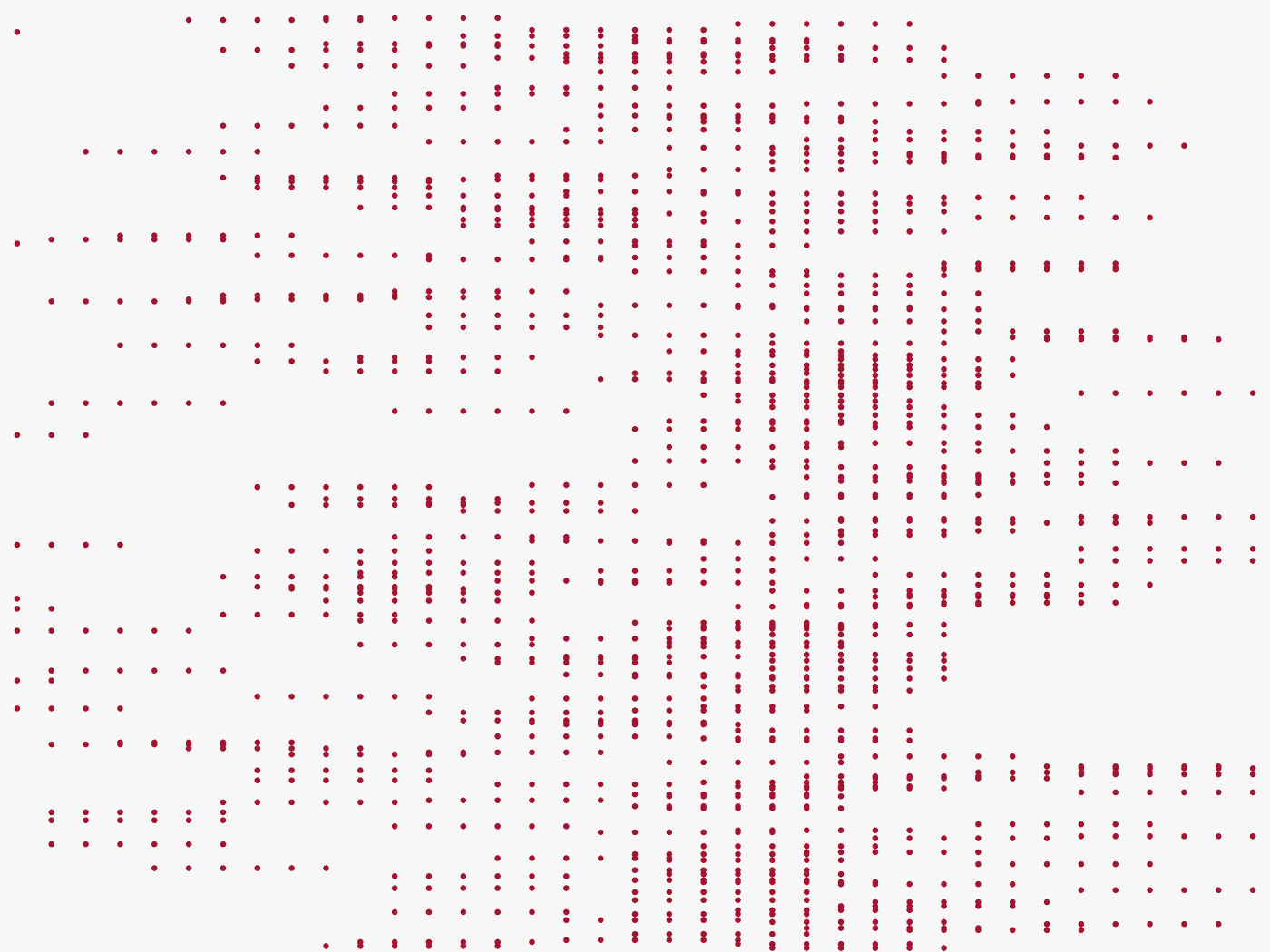}}
	\caption{Examples of reconstructed signals from a sample of $20\%$
		of the node data, using the described classifier with
		$\tau = 0.5$. }
	\label{fig:ReconstructedSignals}
\end{figure*}

The superiority of our classifier compared to the benchmarks also becomes evident when looking at the reconstructed signals. Fig.~\ref{fig:ReconstructedSignals} shows an example with the original signal shown in Fig.~\ref{ex:ExampleSignal}. We observe that the sparsity in the Laplacian/adjacency matrix-based reconstructions appears random, whereas for our novel DAG-based method the signals have structure as often values along time steps (horizontally) tend to stay constant which captures the causal nature of the infection signal.

\section{Conclusion}
\label{sec:Conclusion}%

We presented a novel linear SP framework including shift, convolution, and Fourier analysis for signals on DAGs, or, equivalently, posets. Doing so is significant both theoretically and practically. On the fundamental side we fill a blind spot in graph SP, for which digraphs are problematic and DAGs are a worst case. For applications, DAGs are the natural index domain for causal data, in which each data point causally depends on the values of predecessors. We argue that if this causal dependency is linear, and with coefficients obtained by a suitable transitive closure of the DAG, the signal and causes can be viewed as a Fourier pair. If the linear relation is not causal our proposed Fourier analysis is still mathematically sound but the spectrum cannot be interpreted as causes.

Importantly, our framework allows for edge-weighted DAGs, and, different from other non-Euclidean SP frameworks, there is a degree of freedom in their interpretation and thus the transitive closure needed to obtain a Fourier basis. 

One particular application domain are DAGs obtained from dynamic graphs evolving over discrete time. We show an example of learning infection signals on such a graph from few samples by assuming Fourier-sparsity. 

Overall, our work leverages but also extends the classical theory of Moebius inversion to define a new notion of Fourier analysis for use in signal processing and learning.

\section*{Acknowledgements}

We thank Panagiotis Misiakos for the insight on the relationship between structural equation models and our novel Fourier analysis for DAGs.

\bibliographystyle{IEEEbib}
\bibliography{Literature}


\appendix


\mypar{Proof of Theorem~\ref{th:wmi}}
The proof of the weighted Moebius inversion \eqref{eq:WeightedMoebiusInversion} generalizes the Moebius inversion in~\cite{Rota:64} and is similar to the proof of~\cite[Lemma~2.2.1]{Hall:1998a}.

First we show that 
\begin{equation*}
	\sum_{z \leq y \leq x} w_{x,y} \mu_{w}(z,y)
	=
	\begin{cases}
		1 & \text{if } z = x, \\
		0 & \text{otherwise}.
	\end{cases}
\end{equation*}
The first case holds since $w_{x,x} = 1$. For the second case,
\begin{equation*}
	\begin{split}
		\sum_{z \leq y \leq x} w_{x,y} \mu_{w}(z,y)
		&=  \sum_{z \leq y < x} w_{x,y} \mu_{w}(z,y) +
		\mu_w(z,x)\\
		&= \sum_{z \leq y < x} w_{x,y} \mu_{w}(z,y) \\
		&\qquad - \sum_{z \leq y < x} w_{x,y} \mu_w(z,y) \\
		&= 0,            
	\end{split}
\end{equation*}
where we used the definition of $\mu_w$ in Theorem~\ref{th:wmi}.

With this we can write
\begin{equation*}
	\begin{split}
		s_x
		&= \sum_{z \leq x} \sum_{z \leq y \leq x} w_{x,y} \mu_{w}(z,y) s_z,  \\
		&= \sum_{y \leq x} w_{x,y} \sum_{z \leq y} \mu_w(z,y) s_z.    
	\end{split}
\end{equation*}
Thus, the formula for $c_y$ in Theorem~\ref{th:wmi} implies the formula for $s_x$, and since $W$ is invertible, the reverse holds as well.

\mypar{Proof of Theorem~\ref{thm:PropsTotalVariation}}
Using \eqref{eq:matdiag}, we get
\begin{equation*}
	T_q \coord{f}^y =
	\begin{cases}
		\coord{f}^y & \text{if }y \leq q, \\
		0 & \text{otherwise},
	\end{cases}
\end{equation*}
which also holds after normalization. It follows $\TV_q(\coord{f}^y) = \norm{\coord{f}^y - T_q\coord{f}^y}_2 = 1$ if $y\not\leq q$ and $=0$ otherwise, which yields \eqref{eq:TotalVariationOfNormalizedEV} and
\eqref{eq:SumTotalVariationOfNormalizedEV}.

For the isomorphic partial ordering assume first $x \leq y$ for $x,y \in\vrt$. Then
$y \leq q$ implies $x \leq q$, i.e., $x \not\leq q$
implies $y \not\leq q$. It follows $\TV_q(\coord{f}^x) \leq \TV_q(\coord{f}^y)$.

For the reverse assume $\TV(f^x) \leq \TV(f^y)$, i.e., $\TV_q(\coord{f}^x) \leq \TV_q(\coord{f}^y)$ for all $q\in\vrt$. It follows that $x \not\leq q$ implies $y \not\leq q$, i.e., that $y \leq q$ implies $x \leq q$. Setting $q = y$ yields the result.

\end{document}